\documentclass[longauth]{aa}

\usepackage{graphicx}
\usepackage{txfonts,booktabs}
\usepackage[colorlinks=true,citecolor=blue,linkcolor=blue]{hyperref}
\usepackage{natbib}
\bibpunct{(}{)}{;}{a}{}{,} 
\usepackage{bm}
\usepackage{upgreek}
\usepackage[dvipsnames]{xcolor}
\usepackage{amsmath}    
\usepackage{threeparttable}
\usepackage[modulo, switch]{lineno}
\usepackage{orcidlink}
\usepackage{longtable}
\usepackage{array}

\newcommand{\kmps}{km s$^{-1}$}

\newcommand{\fsie}{f_{\rm SIE}}

\newcommand{\as}{$^{\prime \prime}$}

\begin{document} 


    \title{TDCOSMO}
    
    \subtitle{XXI. Accurate stellar velocity dispersions of the SL2S lens sample and the fundamental plane of the lensing mass
    }
    
    \titlerunning{Accurate stellar velocity dispersions of the SL2S sample}
    \authorrunning{P.~Mozumdar, S.~Knabel et al.}
    
    \author{
        Pritom Mozumdar\orcidlink{0000-0002-8593-7243},\inst{1} \thanks{These authors contributed equally and are considered co-first authors. \email{pmozumdar@astro.ucla.edu}}   	
        Shawn Knabel,\inst{1}$^{\star}$  	
        Tommaso~Treu\orcidlink{0000-0002-8460-0390},\inst{1}
        Alessandro~Sonnenfeld,\inst{2, 3, 4}
        Anowar~J.~Shajib\orcidlink{0000-0002-5558-888X},\inst{5, 6, 7}
        Michele~Cappellari,\inst{8}
        Carlo~Nipoti\orcidlink{0000-0003-3121-6616}\inst{9}
    }
    
    \institute{
    Department of Physics and Astronomy, University of California, Los Angeles, CA 90095, USA; \email{pmozumdar@astro.ucla.edu}
    \and
    Department of Astronomy, School of Physics and Astronomy, Shanghai Jiao Tong University, Shanghai 200240, China
    \and
    Shanghai Key Laboratory for Particle Physics and Cosmology, Shanghai Jiao Tong University, Shanghai 200240, China
    \and
    Key Laboratory for Particle Physics, Astrophysics and Cosmology, Ministry of Education, Shanghai Jiao Tong University, Shanghai 200240, China
    \and
    Department  of  Astronomy  \&  Astrophysics,  University  of Chicago, Chicago, IL 60637, USA
    \and
    Kavli Institute for Cosmological Physics, University of Chicago, Chicago, IL 60637, USA
    \and
    Center for Astronomy, Space Science and Astrophysics, Independent University, Bangladesh, Dhaka 1229, Bangladesh
    \and
    Sub-Department of Astrophysics, Department of Physics, University of Oxford, Denys Wilkinson Building, Keble Road, Oxford, OX1 3RH, UK
    \and
     Dipartimento di Fisica e Astronomia ``Augusto Righi'', Alma Mater Studiorum - Università di Bologna, via Gobetti 93/2, 40129, Bologna, Italy
    }
             
   \date{Received xxx, xxxx; accepted xxx, xxxx}

 
  \abstract
  {We reanalyzed spectra that were taken as part of the SL2S lens galaxy survey with the goal to obtain the stellar velocity dispersion with a precision and accuracy sufficient for time-delay cosmography. In order to achieve this goal, we imposed stringent cuts on the signal-to-noise ratio (S/N), and employed recently developed methods to mitigate and quantify residual systematic errors that are transferred from template libraries and fitting process. We also quantified the covariance across the sample. For galaxy spectra with S/N $>20/$\AA, our new measurements have an average random uncertainty of 3-4\%, an average systematic uncertainty of 2\%, and a covariance across the sample of 1\%. We find a negligible covariance between spectra taken with different instruments. The systematic uncertainty and covariance need to be included when the sample is used as an external dataset in time-delay cosmography. We revisited empirical scaling relations of lens galaxies based on the improved kinematics. We show that the SL2S sample, the TDCOSMO time-delay lens sample, and the lower-redshift SLACS sample follow the same correlation of the effective radius, stellar velocity dispersion, and lensing mass, known as the lensing-mass fundamental plane, as the previously  derived correlation that assumed isothermal mass profiles for the deflectors. We also derived for the first time the lensing-mass fundamental plane assuming free power-law mass density profiles, and we show that the three samples also follow the same correlation. This is consistent with a scenario in which massive galaxies evolve by growing their radii and mass, but stay within the plane.}
  

  \maketitle
%
\section{Introduction}

The Hubble tension is arguably one of the most pressing open problems in cosmology \citep[e.g.,][and references therein.]{Cosmoverse}. The tension between the Hubble constant H$_0$ inferred from early- and late-universe probes has surpassed the canonical 5$\sigma$ threshold for a statistically significant discrepancy. If the tension is not due to as yet unidentified systematic errors, it implies that new physics is required beyond the canonical flat $\Lambda$CDM. 

In order to quantify whether the tension is caused by systematic errors, methods for measuring H$_0$ independently of the traditional local distance ladder and the sound horizon at recombination are especially valuable. Among these, the so-called time-delay cosmography stands out \citep{Refsdal64,Treu16,Treu22,Shajib24}. It is completely independent of all other methods, it measures distances that are large enough to not be affected by local over or underdensities and peculiar motions, but are small enough to be regarded as a late-time probe.

The most advanced time-delay cosmography relies on measurements of the stellar velocity dispersion of galaxies that act as deflectors \citep{treu02b,koopmans03,shajib18,Yildirim20,birrer20,Shajib23} to break the mass-sheet degeneracy \citep{Falco85}. The requirements for measurements of the stellar velocity dispersion for time-delay cosmography are much more stringent than for galaxy evolution studies, however. The relative error on H$_0$ is approximately twice the relative error on the stellar velocity dispersion, and the systematic errors and covariances between measurements therefore need to be controlled at the percent level or better in order to achieve the 2\% precision and accuracy required to distinguish between the Planck \citep{Planck_collab_2018} and `Supernovae, HO, for the Equation of State of Dark energy' (SH0ES) measurements \citep{Riess22}.  Motivated by this requirement, our team carried out a detailed investigation of the systematics of the stellar velocity dispersion in \citet{Knabel_Mozumdar_2025} and showed that a subpercent accuracy can be reached with curated stellar template libraries and methods when spectra with a sufficient S/N and resolution are available.

In this paper, we reanalyze the spectra that were taken a decade ago as part of the `Strong Lensing Legacy Survey' (SL2S) \citep{Ruff11,Gavazzi12, Sonnenfeld_2013a,Sonnenfeld13b,Sonnenfeld14,Sonnenfeld15} with the goal of implementing the recent improvements in the derivation of stellar kinematics to obtain accurate measurements and estimates of the systematic errors and covariance between galaxies. SL2S is a very valuable external dataset for cosmography because it is well matched to the time-delay lenses in terms of redshift distribution and provides information that is helpful for breaking the mass-sheet degeneracy \citep{B+T21}, as demonstrated in a recent milestone paper by the TDCOSMO collaboration \citep{Milestone25}. While the precision of 5-10\% obtained by \citet{Sonnenfeld13b} was sufficient to meet the main goal of the SL2S papers, that is, to characterize the cosmic evolution of luminous and dark matter in early-type galaxies \citep{William_2025_dinos_2}, additional effort and more modern techniques are required to use them as external datasets for cosmography, however. 

We also took advantage of the new measurements to investigate whether the SL2S sample galaxies obey the same lensing-mass fundamental plane as the lower-redshift `Sloan Lens ACS Survey' (SLACS) sample \citep{Bolton07,Bolton08b,Auger_2010}. The lensing-mass fundamental plane is an empirical correlation similar to the traditional fundamental plane \citep{Dressler1987,DD1987}. However, it replaces the surface brightness component of the traditional fundamental plane with the total surface mass density obtained via the lens model. The lensing-mass fundamental plane is tighter than the traditional fundamental plane and closer to the simple virial relation, which indicates a high degree of similarity between the mass distribution and kinematic structure of massive elliptical galaxies \citep{NTB08}. We used the canonical lensing-mass fundamental plane based on the assumption of an isothermal mass density profile, and we introduce for the first time a new plane that is based on the lensing-only slope measurements to extrapolate the surface mass density to the standard circular aperture with a radius equal to half the effective radius. We show that the SL2S and SLACS lenses follow the same planes as the time-delay lenses of the TDCOSMO sample \citep[]{Milestone25, Millon_2020_TDCOSMO_I,birrer20}. This finding is consistent with the assumption that external datasets have the same mass-structure as the time-delay lenses that are required to obtain joint constraints \citep{birrer20,B+T21}.

In Sect.~\ref{sec:data} we summarize the SL2S spectroscopic data. In Sect.~\ref{sec:kinematics} we describe our method. In Sect.~\ref{sec:new} we present our new measurements of the stellar velocity dispersions and their uncertainties. In Sect.~\ref{sec:comparison} we compare our new measurements with the previously published ones, and we study the lensing-mass fundamental plane of the SL2S sample in comparison with SLACS and TDCOSMO.  In Sect.~\ref{sec:summary} we summarize and discuss our results. When necessary, we adopt a flat $\Lambda$ cold dark matter cosmology, with $\Omega_{\rm m}=0.3$, $\Omega_{\Lambda}=0.7$, and $h=0.7$.

\section{SL2S spectroscopic data}
\label{sec:data}

The spectroscopy for 58 SL2S lens candidates was obtained over several years with the goals of confirming the lensing hypothesis, securing source and deflector redshift, and measuring the stellar velocity dispersion. Five of these candidates are not lens systems, and the other 53 were analyzed by \citet{Sonnenfeld13b}. In this work, we excluded 6 of these 53 systems because the instrumental resolution was insufficient and/or the S/N was very low, as described below in more detail. When multiple spectra were available for a galaxy from the same or different instruments, we used the spectrum with the highest S/N as our fiducial spectrum. We also report the results from the second-best spectra for completeness, and we use them to perform a test of systematic errors. 

In this section, we briefly describe the three telescope/instrument setups that yielded the spectra of the galaxies in this sample. For specific details of the observations, such as observing date, slit width, total exposure, and seeing, we refer to Table~2 in \citet{Sonnenfeld13b}. The original HST images of the SL2S lens systems were presented by \citet{Sonnenfeld_2013a} and were later updated with newly acquired observations by \citet{T24} and \citet{William_2025_dinos_2}.

\subsection{Keck-LRIS}

\begin{figure*}
\includegraphics[width=\textwidth]{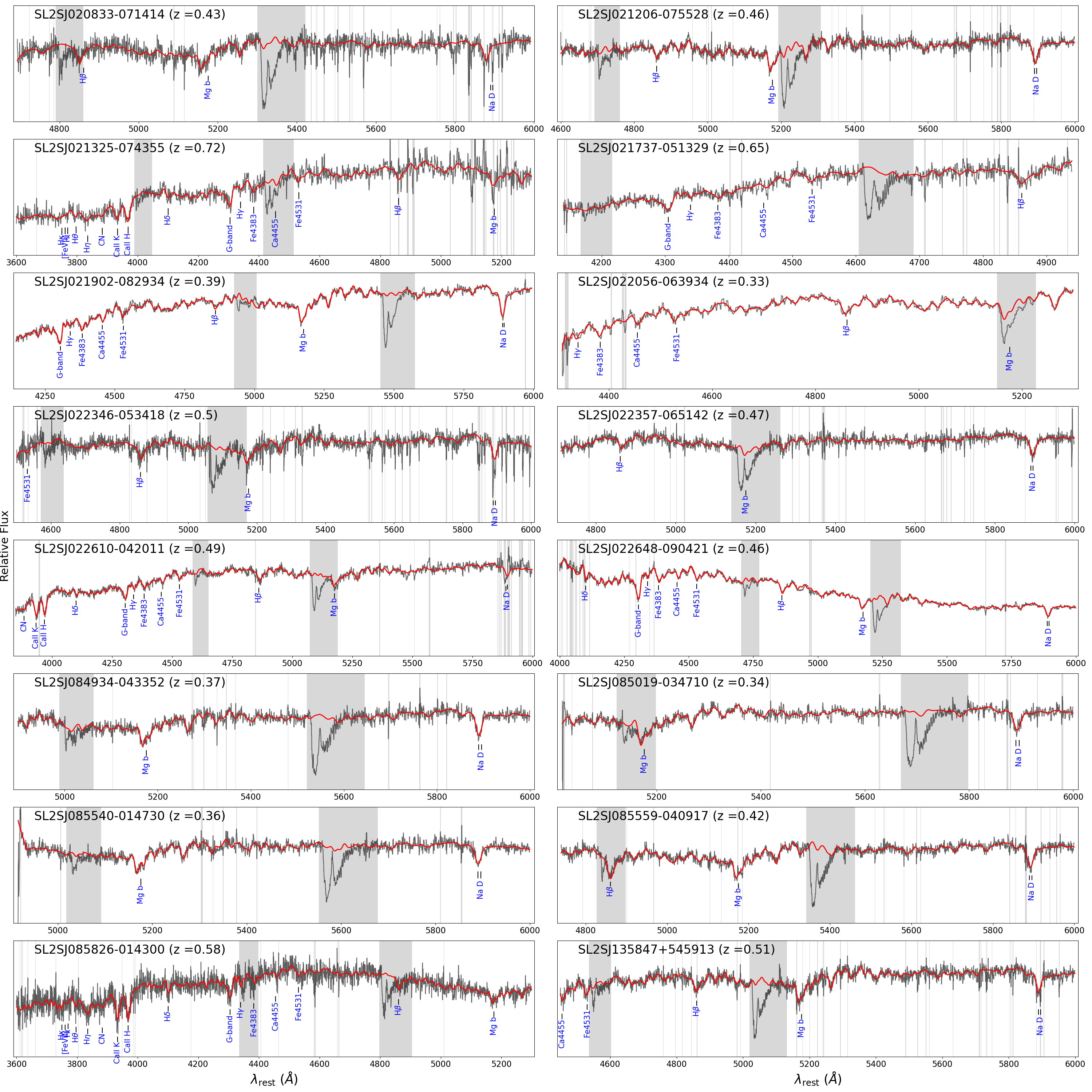}
	\caption{\label{fig:LRIS_galaxy_1}
	  Spectra of the 16 of the 31 lens galaxies that were observed with LRIS on the Keck Telescope. In each panel, the black line shows the observed spectrum, and the red line shows the best-fit model from pPXF. The gray regions mark the wavelength range that we excluded from the fitting and typically these regions correspond to atmospheric absorption lines. Each panel also indicates the name of the system, the lens galaxy redshift, and prominent absorption features in the spectra.}
\end{figure*}

\begin{figure*}
\includegraphics[width=\textwidth]{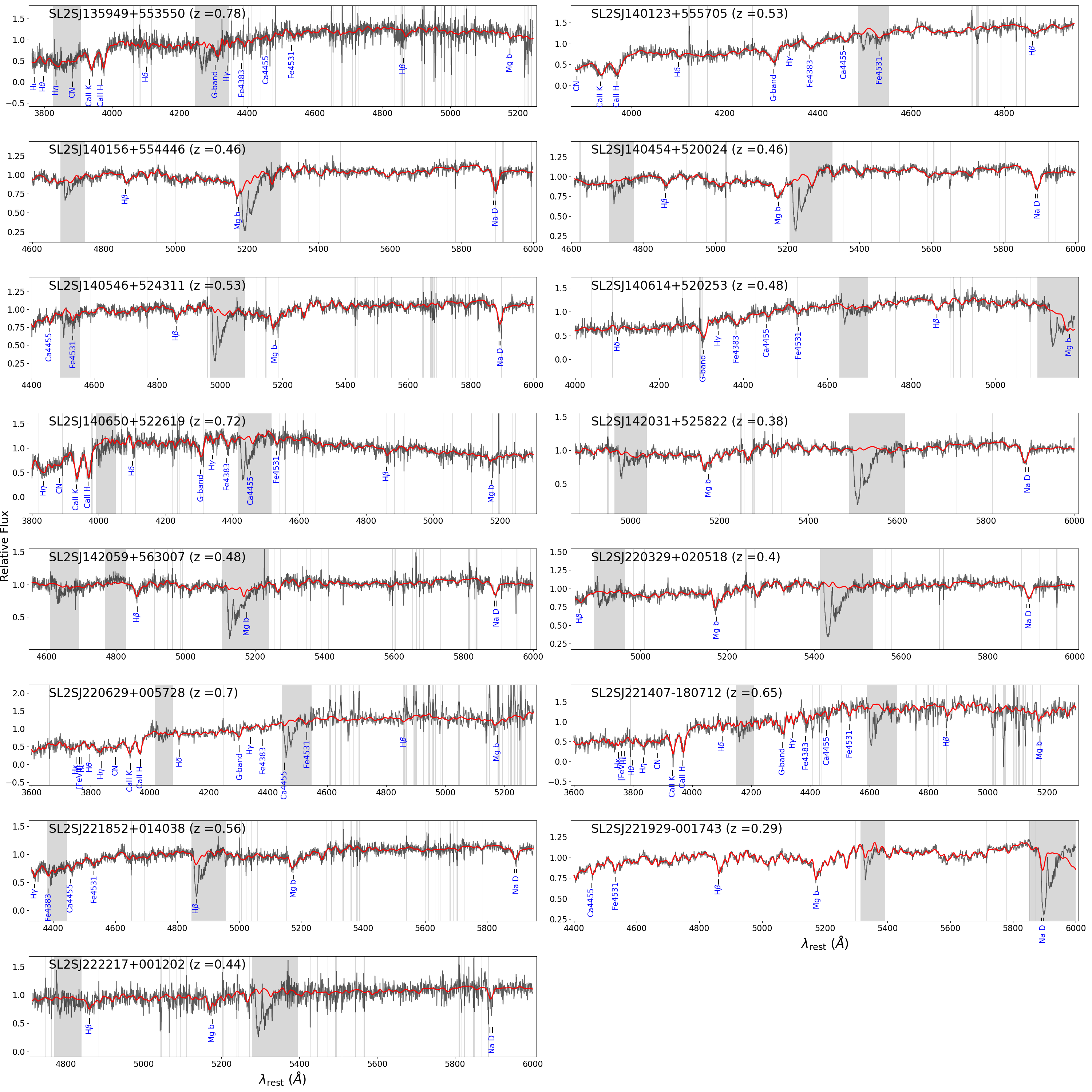}
	\caption{\label{fig:LRIS_galaxy_2}
	Spectra of the remaining 15 of the 31 galaxies observed with LRIS. The line styles of the figure are same as Fig. \ref{fig:LRIS_galaxy_1}.}
\end{figure*}

Thirty-one of the 47 lens galaxies in this sample were observed with the Low-Resolution Imaging Spectrometer (LRIS; \citealt{Oke95}) on Keck. LRIS consists of two detectors (blue and red) that cover a wavelength range of 3000-10000 \r{A}. The spectral resolution and reciprocal dispersion depend on the configuration of the grism (blue side) and grating (red side). The spectra were collected in the long-slit mode using both arms simultaneously with different exposure times. The observing nights were spread over several years, from 2006 to 2012, with typical seeing conditions (FWHM) ranging from  $\sim$ 0.6 \as\ - 1.3 \as. We only used spectra from the red side as the instrumental resolution for the setup used on the blue side was very low.
The 31 spectra with a sufficient spectral resolution and S/N for determining the stellar velocity dispersions are shown in Figs.~\ref{fig:LRIS_galaxy_1} and~\ref{fig:LRIS_galaxy_2}. 

\subsection{Keck-DEIMOS}

\begin{figure*}
\centering
\includegraphics[width=\textwidth]{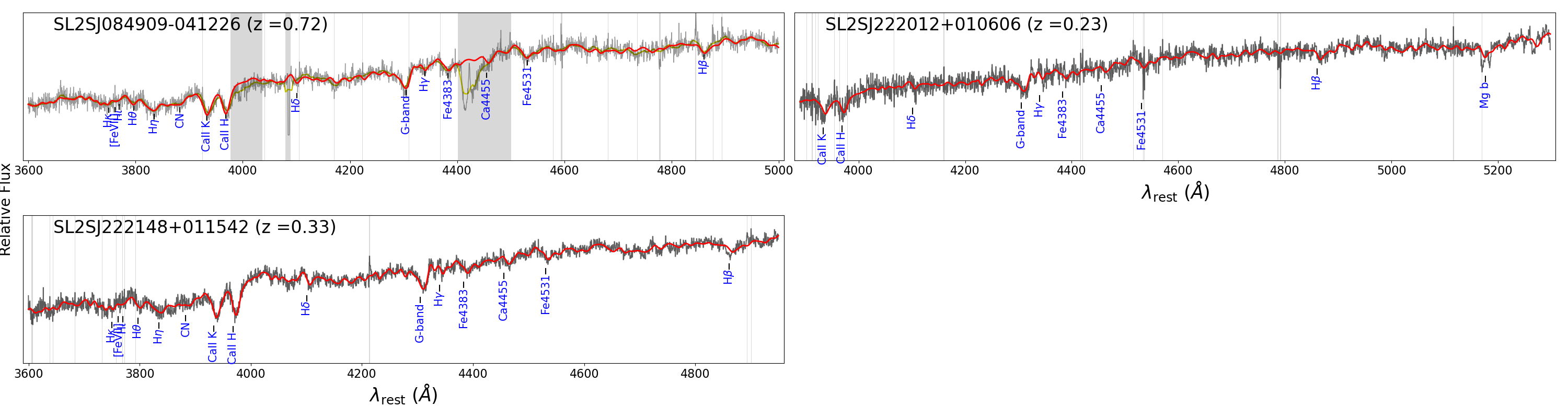}
	\caption{\label{fig:DEIMOS_galaxy}
	Spectra of three lens galaxies from the SL2S sample observed with DEIMOS on the Keck Telescope. In each panel, the black line shows the observed spectrum, and the red line shows the best-fit model from pPXF. The gray regions mark the wavelength range we excluded from the fit.}
\end{figure*}

The data for three systems in this sample were obtained with the `DEep Imaging Multi-Object Spectrograph' \citep[DEIMOS;][]{Faber_2003_DEIMOS} at Keck Observatory. The observations were performed using a long slit with a width of 1\as\ with a 600ZD grating. This setup covers a wavelength range of 4500 - 9500 \AA\ with a reciprocal dispersion of 0.65 \AA. The typical instrumental resolution is about $\sigma_{\rm inst} \sim$ 70 \kmps. The spectra were collected during two observing nights in 2012. The three spectra  with sufficient spectral resolution and S/N ratio for determining the stellar velocity dispersions are shown in Fig.~\ref{fig:DEIMOS_galaxy}. 

\begin{figure*}
\includegraphics[width=\textwidth]{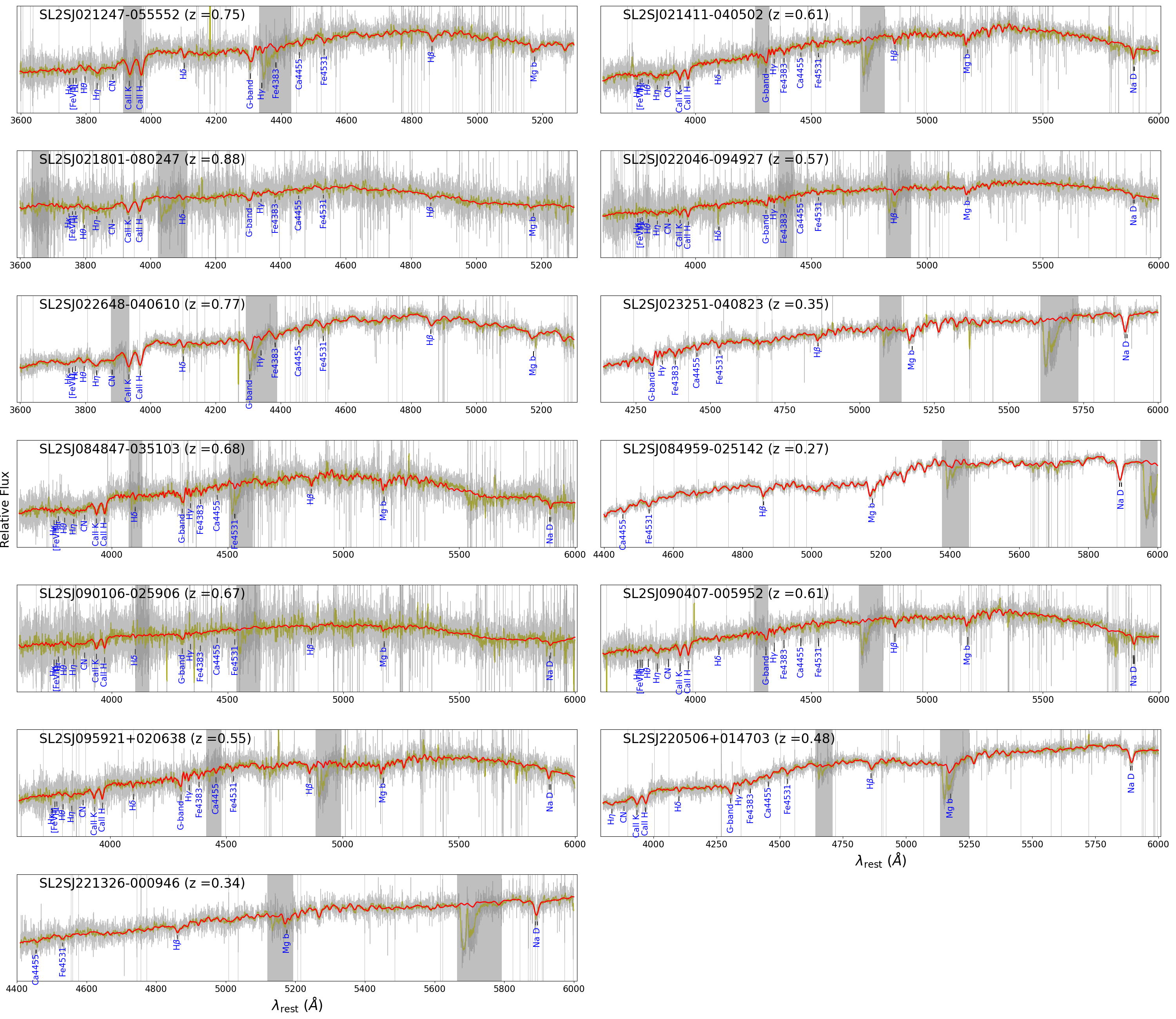}
	\caption{\label{fig:XSL_galaxy}
	Spectra of 13 lens galaxies observed with the Xshooter spectrograph on the VLT. In each panel, the gray line shows the observed spectrum, and the red line presents the best-fit model from pPXF. We also plot the smoothed version (the yellow line) of the observed spectrum. The gray bands mark the wavelength range we excluded from the fit and typically they correspond to the atmospheric absorption lines. Each panel also indicates the name of the system, the lens galaxy redshift, and prominent absorption features in the spectra.}
\end{figure*}

\subsection{VLT-Xshooter}

The Xshooter spectrograph on VLT \citep{Vernet_2011} was used to obtain data for 13 systems. Xshooter consists of three independent cross-dispersed echelle spectrographs in three bands (UVB, VIS, and NIR) and covers a huge wavelength range of 3000 - 25000 \AA. This instrument has a higher resolution than LRIS and DEIMOS. The observations were performed using long slits with a width of 1.0\as\ for the UVB arm, and 0.9\as\ for the VIS and NIR arm. The observing nights were spread over three years from 2010 - 2012. The 13 spectra with a sufficient spectral resolution and S/N for determining the stellar velocity dispersions are shown in Fig.~\ref{fig:XSL_galaxy}.

\section{Method for new kinematic measurements} 
\label{sec:kinematics}

\subsection{Summary of the method}
\label{sec:method}
We applied the recipe proposed by \citet[][hereafter KM25]{Knabel_Mozumdar_2025} to measure the stellar kinematics and minimize and estimate the associated systematic uncertainties. Using mock spectra with different S/N, KM25 assessed the recovery of the true velocity dispersion as well as the resulting uncertainty estimates and showed that the proposed method achieves the desired accuracy on mock samples. A detailed description of the simulation process is provided in the appendix of KM25. First, we measured the instrumental line spread function (LSF) $\sigma_{\text{inst}}$ of the observed galaxy spectra. We also measured the S/N of each spectrum and, using simulations, we determined the S/N required for each specific instrument to keep any bias below the 1\% level. Using the information from KM25, we constructed two clean stellar libraries, Indo-US\footnote{\url{https://noirlab.edu/science/observing-noirlab/observing-kitt-peak/telescope-and-instrument-documentation/cflib}} \citep{Valdes04}, and MILES\footnote{\url{https://miles.iac.es/}} \citep{Sanchez_2006, Falcon-Barroso11}, which were customized for the wavelength range we used in the kinematics fitting, as described in more detail below.

Second, improper flux calibration, variance in continuum strength, and so on may bias the fit significantly when they are not corrected for using additive and multiplicative polynomials. To determine optimal orders of the polynomials and strike a balance between under- and overfitting, we ran a grid of additive and multiplicative polynomials. For low polynomial orders, the fits are generally unstable, leave significant residuals, and result in disagreements between the template libraries that likely arise from an incorrect flux calibration. Above a certain threshold that depends on the wavelength range and on the quality of the flux calibration, the result becomes stable and yields a good agreement between the templates. When the polynomial order is too high, the polynomials begin to modify the shape of the absorption lines, and the fit again becomes unstable. This shows that it is important to choose polynomial orders in the sweet spot, as discussed by KM25. For this dataset, orders of 8 and 2 for additive and multiplicative polynomials, respectively, provide the least systematic scatter between templates, although similar results were found for somewhat lower and higher orders. 

Third, we selected the wavelength range to fit. We intended to use the widest possible wavelength range available for a galaxy spectrum to maximize the information content without compromising the S/N.

Fourth, following the KM25 recipe, we used the Bayesian information criterion (BIC) to estimate the weighted average velocity dispersion ($\bar{\sigma}$) and the associated systematic ($\Delta \bar{\sigma}$) and statistical uncertainties ($\delta \bar{\sigma}$).

An important step in accomplishing this goal was to measure the threshold level for stellar templates to be considered as a significant contribution to the fit. For a threshold level of 1\%, the templates with a weight (w) $0.0<w<0.01$ do not contribute significantly to the fit (improving $\chi^2$) and should not be considered when the number of templates is counted that are used to create the model. We measured the BIC of the fits for the whole sample, the BIC weight of each stellar library, and the BIC-weighted velocity dispersion and uncertainties using the equations provided by KM25. 

\begin{figure*}
\includegraphics[width=\textwidth]{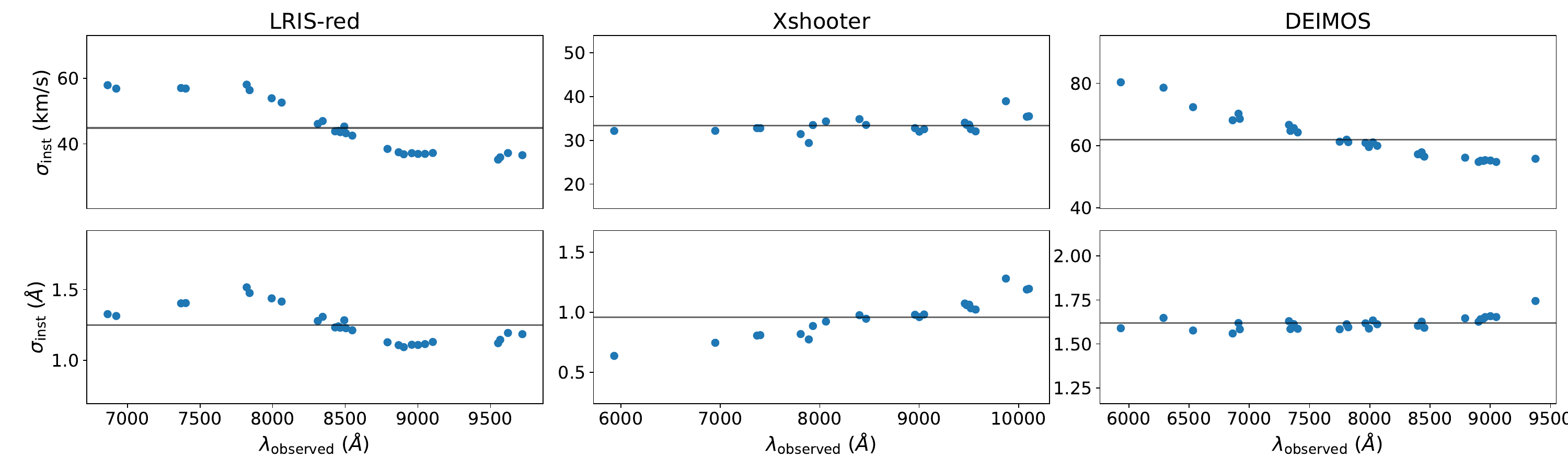}
	\caption{\label{fig:inst_sigma}
	  Example of typical variation in $\sigma_{\text{inst}}$ as a function of wavelength for three instruments: LRIS-red, Xshooter, and DEIMOS. Each plot for each instrument has two panels. The top panel shows $\sigma_{\text{inst}}$ in \kmps\ units, and the bottom panel shows it in \AA\ unit over the observed wavelength. In all panels, the black line shows the average $\sigma_{\text{inst}}$ over the wavelength range.}
\end{figure*}

\subsection{Determining the instrumental resolution and threshold}

We measured the instrumental resolution ($\sigma_{\text{inst}}$) or LSF for each galaxy using its respective variance spectrum. To do this, we fit Gaussian profiles plus a constant (for the background noise) to the isolated and unblended sky lines at different wavelengths. The second moment of the fitted Gaussian profile was considered $\sigma_{\text{inst}}$ at that wavelength. The $\sigma_{\text{inst}}$ was first measured in \AA\ and then converted into \kmps\ using $\sigma_{\text{inst}} = \frac{\Delta \lambda c}{\lambda}$. In Fig.~\ref{fig:inst_sigma} we show the typical wavelength dependence of $\sigma_{\text{inst}}$ for three instruments: LRIS-red, Xshooter, and DEIMOS. As $\sigma_{\text{inst}}$ for all three instruments does not change significantly over the wavelength range we fit, we used the respective average $\sigma_{\text{inst}}$ within the fitted wavelength range of each galaxy for kinematics measurements. We report the average $\sigma_{\text{inst}}$ of each galaxy in Table~\ref{table:veldis_measurements}. The associated uncertainties on $\sigma_{\text{inst}}$ are the sample standard deviations of the measured $\sigma_{\text{inst}}$ at different wavelengths.

We imposed a criterion to ensure that any bias due to imprecise measurement of $\sigma_{\text{inst}}$ was lower than 1\%. Following KM25, this required
\begin{equation}
\frac{\delta \sigma}{\sigma}\approx \frac{\delta \sigma_{\text{inst}}}{\sigma_{\text{inst}}}\left(\frac{\sigma_{\text{inst}}}{\sigma}\right)^2 \lesssim 0.01.
\label{eq:sigmas}
\end{equation}
Thus, depending on the measured $\sigma$, we only accepted instrumental setups for which the criterion was satisfied. We found that all 47 galaxies listed in Table \ref{table:veldis_measurements} satisfied this requirement.

\subsection{Determining the S/N threshold}
It is well known that spectra with an insufficient S/N can result in biased kinematics \citep[e.g.,][]{Treu01}. To determine the S/N required for the unbiased measurements below 1\%, we performed numerical simulations with each specific instrumental setup (resolution, sampling, and so on) used to collect the galaxy spectra of the SL2S sample. We conducted simulations for five setups: LRIS-blue, LRIS-red low-resolution, LRIS-red high-resolution, DEIMOS, and Xshooter. We started by creating mock galaxy spectra using G/K-type stars from the Indo-US and MILES stellar library. We resampled the selected template spectra linearly as per the reciprocal dispersion of a particular instrument setup and convolved the spectra with a Gaussian kernel using pPXF. The $\sigma$ value of the Gaussian kernel was determined by the velocity dispersion to be injected and the instrumental resolution, $\sigma_{\text{inst}}$, of the setup. Then, we added noise to the convolved spectra to mimic a particular S/N. The reciprocal dispersions we used for these setups were 1.45, 1.85, 0.8, 0.65, and 0.2 \AA\ respectively. Similarly, the instrumental resolutions applied on the mock galaxy spectra for the setups were $\sigma_{\text{inst}} =$ 200, 200, 80, 70, and 35 \kmps\ respectively. The chosen reciprocal dispersions and $\sigma_{\text{inst}} $ are representative of the actual SL2S data collected by each setup. For a range of velocity dispersion values ( $\sigma_{\text{input}}$ = 150, 200, 250, and 300 \kmps) and S/N (10, 15, 20, 25, 30, 35, and 40), we created 1000 mock galaxy spectra and recovered the input velocity dispersion using the clean Indo-Us stellar library and the pPXF package\footnote{\url{https://pypi.org/project/ppxf/}} \citep{Cappellari_2017_ppxf,LEGAC_ppxf_Cappellari_2023}. 

Based on these simulations, we found that for LRIS-red high-resolution, DEIMOS, and Xshooter, a S/N threshold of 15 per \AA\ was sufficient to obtain an unbiased estimate below 1\%. The spectral resolution of the LRIS-blue and LRIS-red low-resolution setups is not sufficient to guarantee a bias below 1\% even at a high S/N and velocity dispersion, however. Additionally, as the $\sigma_{\text{inst}}$ of these two setups are comparable to the velocity dispersions of the galaxies to be measured ($\sim$ 200 \kmps), we decided to exclude the spectra from these two setups in this paper. Fortunately, we only had to discard two systems from the sample because of this constraint. In all other cases, complementary data were available from the LRIS-red high-resolution setup.

\subsection{Measuring the S/N of the spectra}

We measured the S/N per \AA\ of all available spectra from the three accepted instrument setups described in the last section. We used three different rest frame wavelength intervals, 4000--4400, 4800--5200, and 5400--5800 \AA, to determine the average S/N per pixel of the spectra and divided it by the square root of the number of \AA\ per pixel to obtain the average S/N per \AA. We selected these three wavelength regions as they have relatively uniform S/N per pixel. When a spectrum did not contain all three regions, we used whatever wavelength ranges from these three regions were available. We excluded pixels covering the atmospheric A and B lines if present in these three regions while measuring the average S/N. We found 30 galaxies with spectra that had an S/N per \AA\ $\geq$ 15. We still measured and report the kinematics and associated uncertainties of the other galaxies, however, because they might be used in projects where a bias above 1\% is acceptable. The S/N of the fitted galaxy spectra is listed in Table~\ref{table:veldis_measurements}.

\subsection{Selection of clean stellar libraries}

KM25 extensively discussed the importance of using clean stellar templates and showed that the usage of defective templates can introduce bias in the kinematics measurements. Therefore, we created two clean stellar libraries, Indo-Us and MILES, using the list of clean stars provided by KM25. Some of the stars in these lists also contain defects in several wavelength regions, and KM25 further flagged these stars with the information of the problematic wavelength range. We checked for the presence of these defective regions in the clean templates within the wavelength range of 3600-6000 \AA. When defects were present, we also excluded these spectra from the clean library. We used the clean Indo-US and MILES library for all the kinematics measurements and systematic tests. The final clean Indo-US and MILES libraries contained 802 and 789 templates, respectively.

We initially also considered using the Xshooter stellar library. More than 50\% of the clean star templates in this library were flagged beyond 5300 \AA\ because of artifacts around the dichroic that separates the UVB from the visible wavelength range, however, while for most galaxies, we had to fit up to 6000 \AA\ at rest-frame. As the clean Xshooter library has to be downsized significantly beyond 5300 \AA, we decided not to use this library in this project.

\subsection{Determination of polynomial orders}

Following the recipe in KM25, we ran a grid of polynomial orders to determine the optimal orders of the additive and multiplicative polynomials that would make the kinematics measurements stable and reduce the systematic scatter in measurements using different template libraries for the whole sample. A similar test was conducted by KM25 using a sample of galaxy spectra that covered the same rest-frame wavelength range and was observed with the same instrument. The galaxy spectra in the SL2S sample contain different rest-frame wavelength ranges and were collected using different instruments, however. Therefore, we also checked whether the optimal polynomial orders depend on the instrument or the fitted wavelength range. We found that higher polynomial orders, such as 7-9 for additive polynomials and 2 for multiplicative polynomials, reduce the systematic scatter regardless of the instrument or wavelength range. We decided to use orders of 8 and 2 for additive and multiplicative polynomials, respectively. As a test, we also computed the additional scatter in stellar velocity dispersion if we marginalized over polynomial orders around the minimal choice, and we found it to be much lower than 0.5\%, which is negligible with respect to that attributable to stellar templates.

\subsection{Wavelength range and masking}

The galaxy spectra of the SL2S sample cover different rest-frame wavelength ranges due to combination of redshift and instrument setup. Regardless, we used the widest possible wavelength range available for a galaxy spectrum without compromising the S/N. The cutoff of the adopted wavelength range on the blue side was imposed by the instrument setup and the stellar template coverage, and on the red side by the redshifts of the galaxies. In Figs. \ref{fig:LRIS_galaxy_1} to\ref{fig:DEIMOS_galaxy}, we show the wavelength range we fit for each galaxy. We excluded some wavelength regions from the fits at two stages. First, we masked the rest-frame wavelength range 6850-6950 \AA\ and 7580-7750 \AA\, corresponding to the atmospheric B and A band, respectively, if present. Next, using an iterative sigma clipping method, we detected wavelength ranges containing emission lines, instrument or extraction related artifacts, cosmic rays, and so on, if any, and also masked them from the fit. In Figs. \ref{fig:LRIS_galaxy_1} - \ref{fig:DEIMOS_galaxy}, these masked regions are shown in gray bands. 

Following KM25, we also tested for systematic bias due to small changes (100-200 \AA) in the fitted wavelength range. For this test, we selected eight galaxies with the longest wavelength coverage from three instruments. Those with Xshooter and DEIMOS spectroscopy have rest-frame wavelength ranges of about 3600 - 6000 \AA\, and those from LRIS-red have a wavelength coverage of about 4400 - 6000 \AA. This long wavelength range is conducive to making small changes on the fitted wavelength range on the blue and red sides while keeping the prominent absorption features such as CaHK, Mgb, or NaD, if present. We trimmed and/or extended the original wavelength range on either end by 100 - 200 \AA\ and fit for kinematics using the same setting as for the fiducial wavelength range. We then calculated the average deviation in the measured velocity dispersions due to wavelength change relative to the respective fiducial velocity dispersion for each galaxy and for all of them together. We found that the sample average deviation is $\overline{\left<\frac{\sigma_{\lambda_k}}{\sigma_0} -1\right>_i} = 0.8\pm0.5$\%. Here, $i$ represents each of the eight galaxy in the sample, $\sigma_0$ is the velocity dispersion from the fiducial wavelength range, and $\sigma_{\lambda_k}$ are the velocity dispersions from the modified wavelength ranges. As the uncertainty from the changes in wavelength is subdominant to the total error budget at the sample level, we ignored this in estimating the final systematic uncertainty.

\begin{figure}
\centering
\includegraphics[width=\columnwidth]{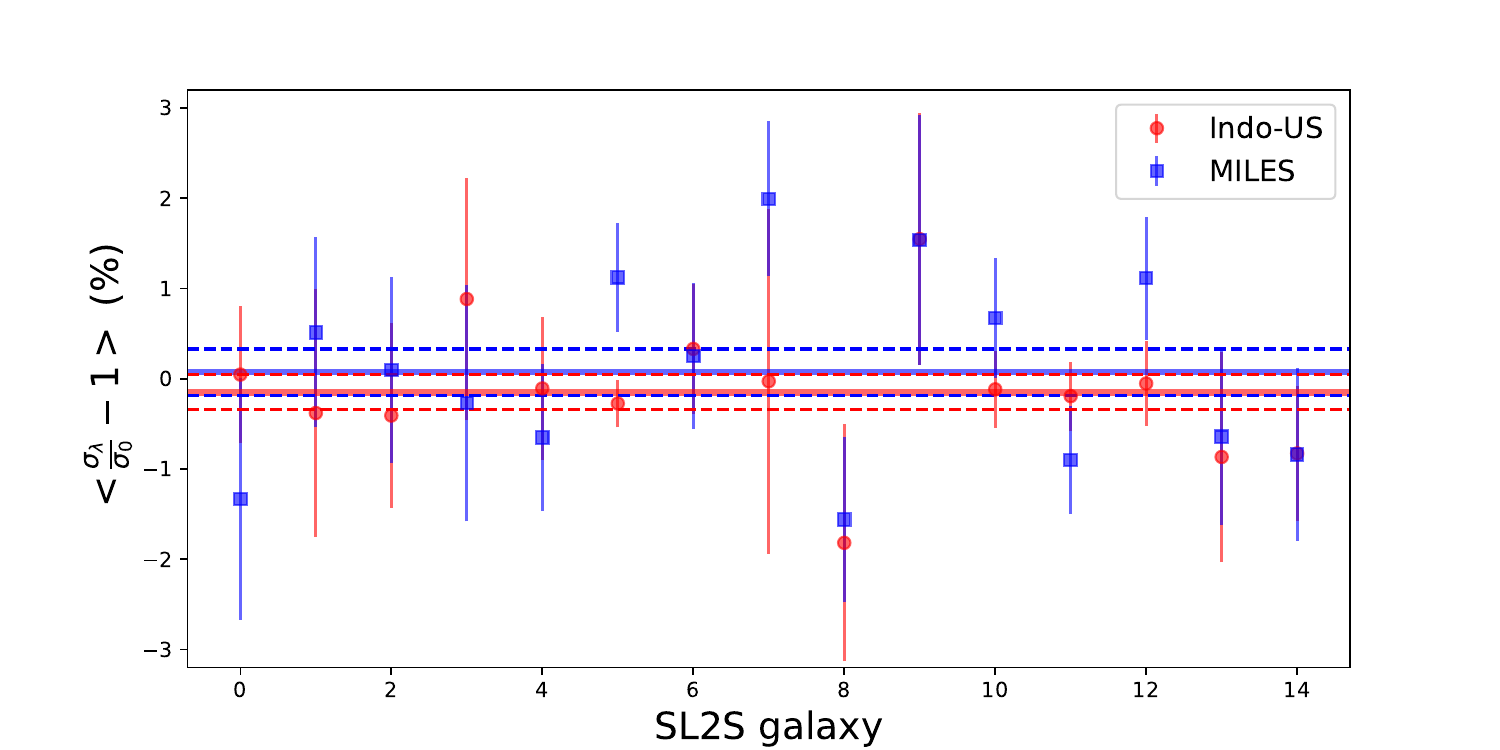}
	\caption{\label{fig:mask}
	For each galaxy with high-quality spectra in the sample (S/N$>20$/\AA), we show the mean and scatter in the value of stellar velocity dispersion when masking one of the main absorption features (CaHK, G-band, MgB, or NaD) with respect to the value obtained for the entire spectrum. The thick horizontal lines represent the sample average, and the dashed lines show the error on the mean.  }
\end{figure}

As an additional test, we repeated the measurement for a subset of galaxies with spectra with an S/N $>20$/\AA, masking at random one of the most prominent absorption features in the spectra (CaHK, Mgb, G-band, or NaD). The results are shown in Fig.~\ref{fig:mask} separately for the Indo-US and MILES library, in order to isolate the effect of the lines from that of the stellar library. The stellar velocity dispersion changes by less than 1\% on average, which demonstrates that the results for data and templates of sufficient quality do not depend on any single feature.

\begin{figure}
\centering
\includegraphics[width=\columnwidth]{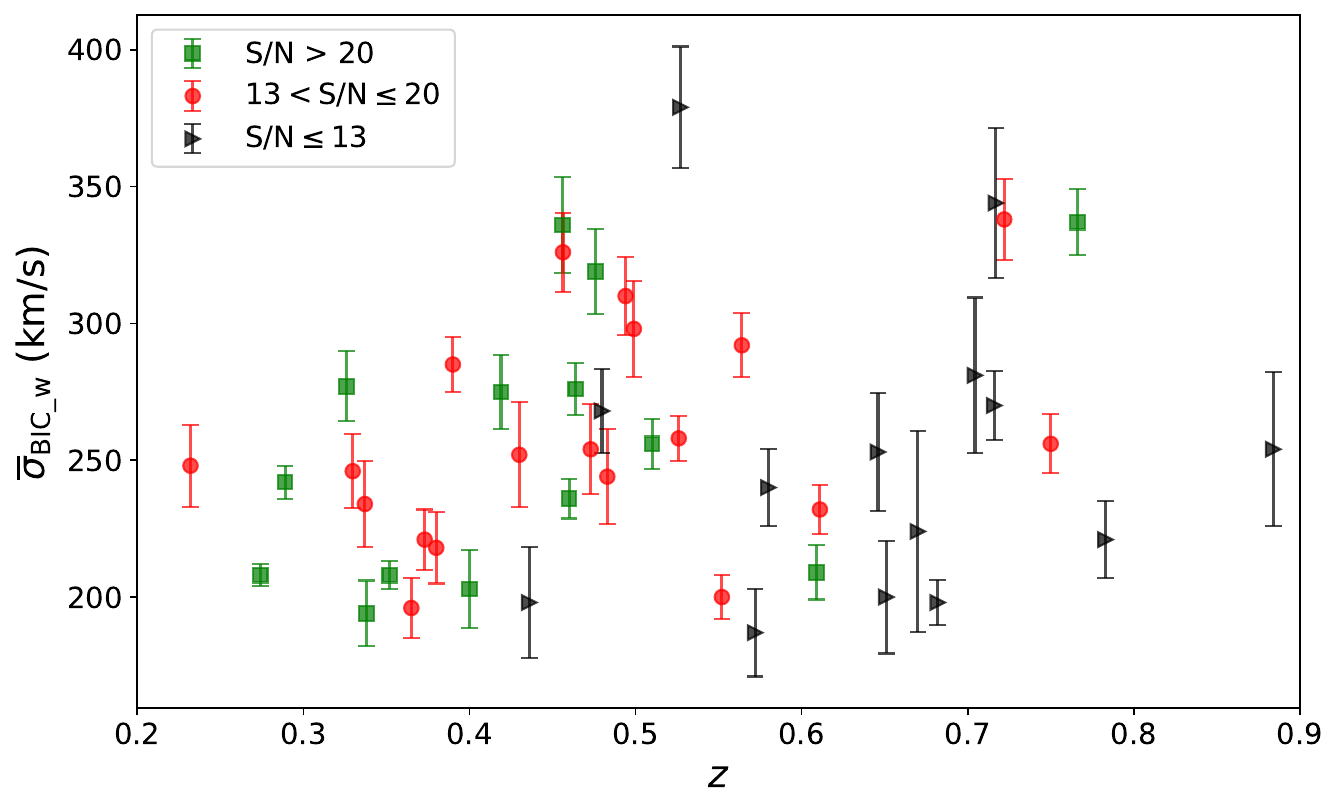}
	\caption{\label{fig:sigma_z}
	BIC-weighted velocity dispersions of the SL2S sample as a function of redshift. The green, red, and black markers show galaxies in three different S/N bins, S/N > 20, 13 < S/N $\leq$ 20, and S/N >20, respectively. The associated error bars show the total uncertainties in the measured velocity dispersions and are dominated by the statistical uncertainties, which are driven by the S/Ns.}
\end{figure}

\subsection{Determining the threshold to count the effective template number}

For the reasons discussed above, we used two clean stellar libraries for the kinematic fitting and combined the measurements and associated uncertainties as a weighted average. The weight assigned to each template library was determined using the BIC for the whole sample. The total BIC and associated weight were calculated following Eqs. (5)-(10) in KM25. One crucial step in measuring BIC is to determine the effective number of templates that provide a significant contribution to the kinematic model. KM25 extensively discussed the issue and how to address it. Briefly, in an example of overfitting the noise, the model could include some templates with very low weights that might improve the fit (lowering $\chi^2$) by a minimal amount. They degrade the BIC significantly by increasing the total number of parameters, however. This effect can lead to incorrect estimates of the weights for the template libraries. Following the procedure in KM25, we first carried out a pPXF fit with all the templates in a clean library, and we then performed successive fits while only using the subset of templates that only contribute above a certain threshold to the initial fit. The threshold is the weight of the templates in defining the model for the initial fit. As expected, for a higher threshold (10-20\%), the BIC is relatively large because there are not enough templates to obtain a good fit. Below $\sim5$\%, however, the $\chi^2$ does not improve enough to compensate for the increased number of degrees of freedom in the BIC. Therefore, we adopted a 5\% threshold to count the effective number of templates for the SL2S sample (as expected, the threshold is similar to the inverse of the S/N of the dataset; see KM25).

\renewcommand{\arraystretch}{1.2}
\begin{table*}
\caption{BIC-weighted uncertainties for the whole SL2S sample and subsamples.}
\begin{center}
\begin{threeparttable}
\resizebox{\textwidth}{!}{
\begin{tabular}{c c c c c c c c c} 
    \hline
Sample    & $<\delta \bar{\sigma}/\bar{\sigma}>$ & $<\Delta \bar{\sigma}/\bar{\sigma}>$ & $\sqrt{<C_{i,j}\bar{\sigma}/\bar{\sigma_i}\bar{\sigma_j}>_{i\neq j}}$ & $<\Delta_B \bar{\sigma}/\bar{\sigma}>$ & $\sqrt{<C_{B,i,j}\bar{\sigma}/\bar{\sigma_i}\bar{\sigma_j}>_{i\neq j}}$ & Indo-US &  MILES & Num. of Obj.\\
\hline
SL2S S/N > 20    &   3.47\%  &  0.66\%   &  0.22\%  &  2.14\%  &  0.72\%   &   0.05 &   0.95 &  15\\ 
SL2S 13 < S/N $\leq$ 20 &  4.60\%  &  1.07\%   &  0.65\%  &  1.70\%   &  1.04\%   &   0.27 &  0.73 &   18 \\
SL2S S/N $\leq$ 13  &   7.40\%  &   1.68\% &  0.90\%   & 3.07\%  & 1.65\%    &   0.81 &    0.19 &  14\\
SL2S All            &   5.05\%  &  1.38\%   & 0.67\%  &  2.22\%  &  1.08\%   &  0.26 &    0.74 &  47\\ 
\hline
SL2S-LRIS S/N > 20    &   3.73\%  &  1.39\%   &  1.07\%  &  2.38\%  &  1.83\%   &   0.78 &   0.22 &  8\\ 
SL2S-LRIS 13 < S/N $\leq$ 20 &  4.68\%  &  1.42\%  &  0.97\%   &  2.09\%   & 1.42\%  &  0.37 &  0.63 &  13 \\
SL2S-LRIS S/N $\leq$ 13  &   6.90\%  &   0.91\% &  0.43\%   & 2.73\%  &  1.30\%    &  0.94 &   0.06 & 10\\
SL2S-LRIS All   &   5.13\%  &  1.63\%   &  1.02\%  &  2.36\%  &  1.48\%   &   0.62 &   0.38 &  31\\ 
\hline
SL2S-Other S/N > 20    &   3.16\%  &  0.003\%   &  0.001\%  &  1.97\%  &  0.45\%   &   0.001 &  0.999 & 7\\ 
SL2S-Other 13 < S/N $\leq$ 20 &  4.38\%  &  0.40\%  &  0.04\%   &  0.71\%   & 0.06\%  &  0.19 &  0.81 & 5 \\
SL2S-Other S/N $\leq$ 13  &  8.68 \%  &   2.49\% &  1.55\%   & 3.69\%  &  2.30\%    &    0.35 &   0.65 &  4\\
SL2S-Other All  &    4.94\%  &   0.10\% &  0.02\%   & 1.96\%  & 0.35\%    &    0.001 &  0.999 & 16\\
\hline
\end{tabular}
}
\end{threeparttable}
\end{center}
\label{table:results}
\tablefoot{We report the BIC-weighted average statistical uncertainty, the average systematic uncertainty (both Bessel corrected and uncorrected), and the average amplitude of the off-diagonal terms of the covariance matrix between elements of the sample (both Bessel corrected and uncorrected) for the whole SL2S sample and subsamples grouped by instrument and S/N. The BIC-determined weights given to the two libraries for each subsample and the number of galaxies in each subsample are also provided.}
\end{table*}

\section{New measurements and uncertainties}
\label{sec:new}

In this section, we present the results we obtained by using the method discussed in the previous section. The fitted models from pPXF are shown in Figs.~\ref{fig:LRIS_galaxy_1} to~\ref{fig:DEIMOS_galaxy}, and the new measurements and uncertainties are summarized in Table~\ref{table:veldis_measurements}. For each galaxy, we report the S/N and instrumental resolution of the spectrum and our best estimate of the stellar velocity dispersion, together with the random and systematic errors. The mean $\bar{\sigma}$, the systematic uncertainty $\Delta \bar{\sigma}$, and the statistical uncertainty $\delta \bar{\sigma}$ of the velocity dispersion of an individual galaxy are defined as in KM25 such that 

\begin{equation}
\bar{\sigma} = \frac{\sum_{k} w_k \sigma_k}{\sum_k w_k}
\label{eq:mean}
\end{equation}
\begin{equation}
(\Delta \bar{\sigma})^2 = \frac{\sum_{k} w_k (\sigma_k-\bar{\sigma})^2}{\sum_k w_k},
\label{eq:sys_error}
\end{equation}
and
\begin{equation}
\delta \bar{\sigma} = \sqrt{\frac{\sum_{k} w_k (\delta \sigma_k)^2}{\sum_k w_k}},
\label{eq:stat_error}
\end{equation}
where the index $k$ runs through the template library, $\delta \sigma_k$ is the statistical uncertainty estimated by pPXF, and $w_k$ is the weight we assigned.

Following KM25, we provide the stellar velocity dispersions and uncertainties as both the BIC-weighted average and equally weighted average of the measurements from two template libraries. The BIC-determined weights were estimated at the sample level to minimize noise in the process, and the samples were grouped by instrument and S/N. The estimated weights for the groups are summarized in Table~\ref{table:results}. The preference for one of the libraries changed depending on the instrument and the S/N. For each galaxy, we used the weight for the group to which it was affiliated to measure the BIC-weighted average values. For example, when a galaxy was observed with LRIS and had an S/N$>$20/\AA, the weights we assigned to the measurements from Indo-US and MILES were 0.78 and 0.22, respectively. In general, the BIC-weighted and equally weighted velocity dispersions are very similar, which is consistent with the good agreement between the template libraries. In Fig. \ref{fig:sigma_z} we show the BIC-weighted velocity dispersions with the associated total uncertainties of the SL2S sample as a function of redshift.\\


When the number of template libraries is small (as in the case considered here), it is prudent to adopt Bessel's correction to obtain the unbiased estimators (see KM25 for a derivation). Thus, the systematic uncertainties were Bessel corrected following
\begin{equation}
(\Delta_B \bar{\sigma})^2 = \frac{ \sum_{k} w_k (\sigma_k-\bar{\sigma})^2}{\sum_k w_k - \sum_k w_k^2 / \sum_k w_k}.
\label{eq:sys_error_Bessel}
\end{equation}
For $k=2$, however, the Bessel correction nullifies the dependence on the weights, and $\Delta_B \bar{\sigma}$ reduces to 
$(\Delta_B \bar{\sigma})^2 = (\frac{\sigma_1 - \sigma_2}{2})^2$, where $\sigma_1$, and $\sigma_2$ are measurements using two template libraries. Thus, the Bessel-corrected systematic uncertainties, $\Delta_B \bar{\sigma}$, for the BIC-weighted and equally weighted velocity dispersions are equal and are reported as a single column in Table \ref{table:veldis_measurements}. 

We also measured the BIC-weighted average statistical and systematic uncertainty, both Bessel corrected and uncorrected, for subsamples grouped by S/N and/or instrumental setup. In addition, we estimated the covariance matrix as follows:

\begin{equation}
C_{ij} \bar{\sigma} = \frac{\sum_{k} w_k (\sigma_{k,i}-\bar{\sigma}_{i})(\sigma_{k,j}-\bar{\sigma}_{j}) }{\sum_k w_k} + \delta_{i,j} \delta \bar{\sigma}_{i}^2,
\label{eq:covariance}
\end{equation}
where the indices $i,j$ identify the elements of the covariance matrix, $\bar{\sigma}_{i,j}$ is the weighted-average velocity dispersion of the element at index $i$ or $j$, and $\delta_{i,j}$ is the Kronecker delta function. As above, the index $k$ runs through the template library, $\delta \bar{\sigma}_i$ is the weighted-average statistical uncertainty, and $w_k$ is the weight we assigned. Along the diagonal, the first term is the systematic error, and the second term is the statistical error. With Bessel correction, the estimator reads 
\begin{equation}
C_{B,ij} \bar{\sigma} = \frac{ \sum_{k} w_{k} (\sigma_{k,i}-\bar{\sigma}_{i})(\sigma_{k,j}-\bar{\sigma}_{j}) }{\sum_k w_k - \sum_k w_k^2 / \sum_k w_k} + \delta_{i,j} \delta \bar{\sigma}_{i}^2.
\label{eq:covariance_Bessel}
\end{equation}

The normalized sample average uncertainties and the average amplitude of the off-diagonal terms of the covariance matrix between elements are summarized in Table~\ref{table:results}. As expected, the random error is driven by the S/N, regardless of the instrumental setup, and it ranges between 3-8\%. The systematic errors also depend on the S/N and instrumental setup, and they lie at a level of 1-2\% without Bessel correction and at about 2-3\% with the correction. This is only the error associated with template choice. Other potential sources of error, such as the wavelength range and polynomial order, were ignored because they are negligible with respect to the contribution from templates.

The average off-diagonal amplitude of the covariance matrix for different subsamples is positive. This is expected as the templates yielding higher (lower) velocity dispersion do so consistently for all galaxies. This quantity ranges around 1-2\% within the same instrumental setup, and it lies at the subpercent level for the different instrumental setups. The LRIS subsamples have slightly higher covariances than the combined Xshooter and DEIMOS samples. 

We found that Indo-US library is preferred by the LRIS data, while MILES is preferred by the other. This emphasizes the importance of using several stellar libraries and calculating BIC-weighted averages. In this case, however, similar results were obtained when the templates were given equal weight, so that the difference in BIC weight had less impact. \\

\begin{figure}
\centering
\includegraphics[width=\columnwidth]{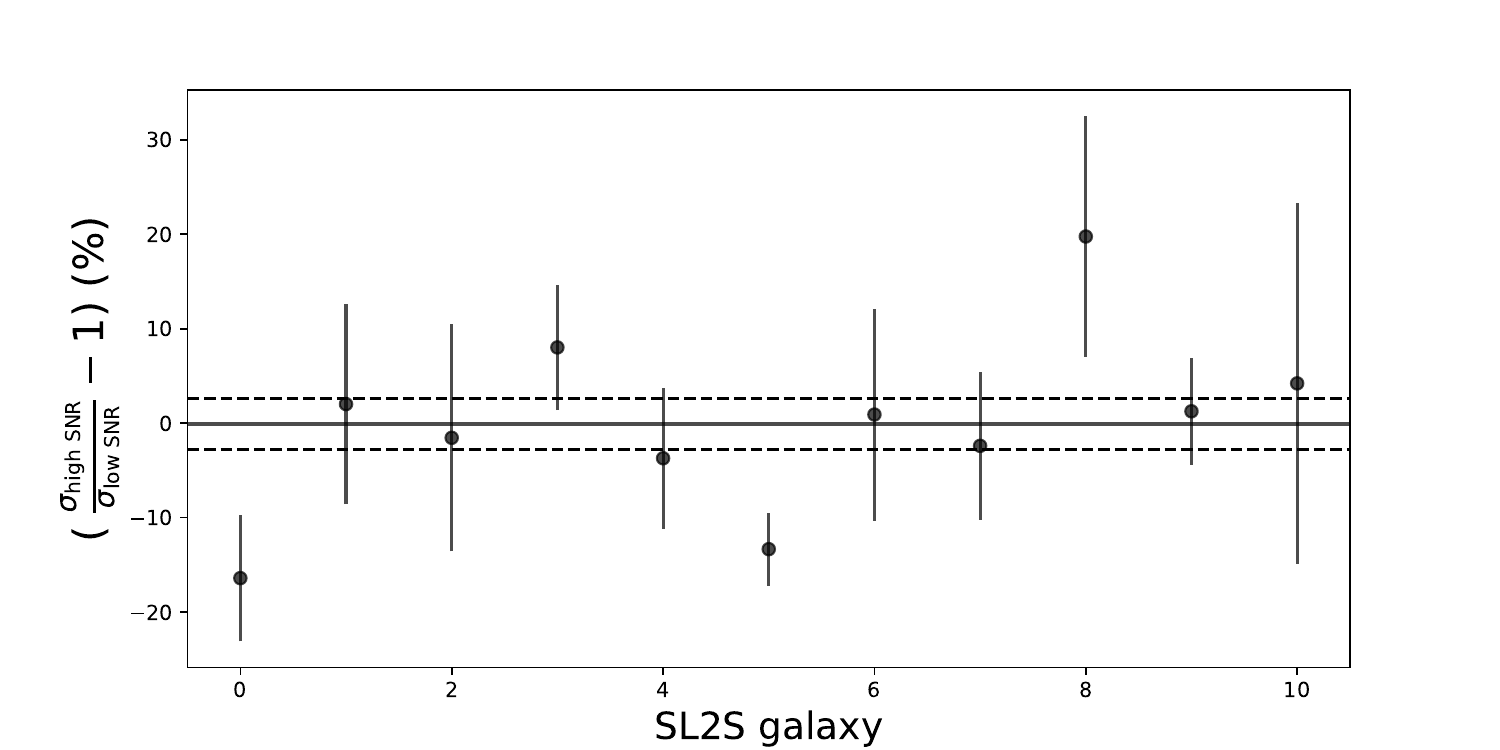}
	\caption{\label{fig:multiplesp}
	Comparison of two measurements from two different spectra for 11 SL2S galaxy. The other spectrum for the same galaxy is either from a different instrument or a different observing night with the same instrument. The $\sigma_{\text{high S/N}}$ is our fiducial measurement, and $\sigma_{\text{low S/N}}$ is the second-best measurement. The solid line shows the mean, and the dashed lines show the 1$\sigma$ uncertainty of the mean.}
\end{figure}

As a consistency test of our error estimates, we compared the stellar velocity dispersion obtained for 11 systems that have multiple spectra with a sufficiently high resolution. The fiducial spectrum was the spectrum with the highest S/N. The other spectra may have significantly lower S/N and may cover wavelength ranges differing by 500-1000 \AA\ from their fiducial counterpart. 
The difference in the S/N quality of the spectra is reflected in the random uncertainties listed in Table~\ref{table:veldis_measurements}. The comparison is shown in Fig.~\ref{fig:multiplesp}, where the error bars represent the sum in quadrature of the random and systematic errors. The sample average of the ratio is $<\frac{\sigma_{\mathrm{high\ S/N}}}{\sigma_{\mathrm{low\ S/N}}} -1> = -0.11\pm 2.70$ \%. Repeated measurements with higher S/Ns are needed to carry out an even more stringent test of the systematics for different instrumental setups.

\section{Comparison with previous work and improved scaling relations}
\label{sec:comparison}

\begin{figure}
\includegraphics[width=\columnwidth]{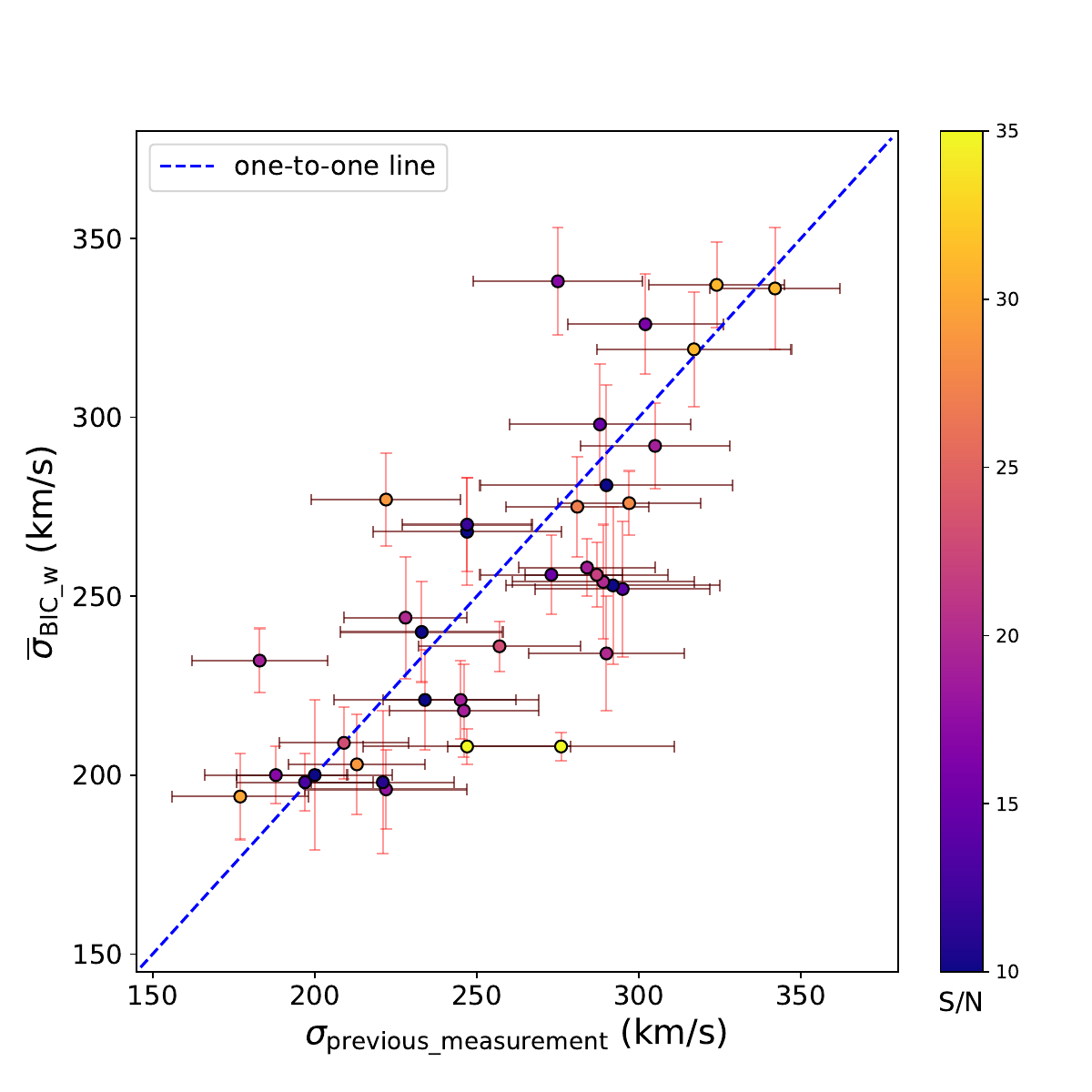}
	\caption{\label{fig:comparison_kinematics}
	Comparison of the velocity dispersions for 37 galaxies as measured in this paper (y-axis) and by \citet[][x-axis]{Sonnenfeld_2013a} based on the same spectra. The galaxies are color-coded by the S/N measured in this work. The method alone changed, i.e., chiefly improvements in the stellar template libraries.}
\end{figure}

In this section, we show that the new measurements are consistent with previous results (Fig.~\ref{fig:comparison_kinematics}), with no average offset. We compared the BIC-weighted velocity dispersion of 37 galaxies with the corresponding measurements by \citet{Sonnenfeld13b}, for which the same spectra were used. For other cases, different spectra were used, or no measurements were reported. We found that for the 37 galaxies, the estimated $<y_i/x_i> = 0.98 \pm 0.02$ where $y_i$ and $x_i$ are new and old measurements for the $i$th galaxy. Thus, the current velocity dispersions are consistent with the previously measured values on average. The reduction in systematic error with respect to the previous analysis is mostly due to three factors. First, stellar template libraries have been improved. Whereas \citet{Sonnenfeld13b} used 7 stars from the Indo-US library, we used hundreds of stars from the cleaned Indo-US and MILES libraries. Second, we did not use the data taken with instrumental setups that have insufficient spectral resolution to meet our target accuracy. Third, we mitigated the effects of continuum and flux calibration uncertainties by modeling them with multiplicative and additive polynomials.

The distribution of the velocity dispersions in the SL2S sample does not vary significantly with redshift (as we showed in Fig.~\ref{fig:sigma_z}), and it peaks at a significantly higher redshift than SLACS ($z=0.1-0.4$). 

We then used the new SL2S measurements to revisit the scaling relations determined by \citet{Auger_2010} for the SLACS sample, taking advantage of the new spectroscopy determinations by \citet{Knabel24} and those presented here. To enable a direct comparison with  \citet{Auger10}, the stellar velocity dispersion was corrected to a fiducial aperture of half the effective radius using the empirical recipes derived by \citet{Knabel24}. We refer to these aperture-corrected velocity dispersions as $\sigma_{1/2}$. We adopted the total uncertainties as the sum in quadrature of the random and systematic errors.

Our goal was to investigate whether there are any differences between the SLACS, SL2S, and TDCOSMO samples \citep[][]{Millon_2020_TDCOSMO_I,birrer20, Milestone25} that might be an indication of structural evolution or differences in terms of these properties. We focused on the SLACS and SL2S galaxies with high S/N spectroscopy from \citet{Knabel24} and this work to minimize the potential impact of systematic errors.  We also focused on the scaling relations that involve lensing quantities because the others might be investigated using much larger samples of nonlens galaxies. The effective radii and Einstein radii were taken from \citet{Auger_2010} for the SLACS sample, from \citet{William_2025_dinos_2} for the SL2S sample, and from \citet{Milestone25} for the TDCOSMO sample. The effective radii were obtained by fitting \citet{Sersic_profile} profiles to the surface brightness of the deflector, jointly with the lensed images\footnote{\citet{deVaucouleurs48} in the case of SLACS-X}.  Noting that the formal errors are usually tiny on these quantities,  we adopted 10\% as our systematic error on the effective radius \citep[e.g.,][sec.~4.1.1]{ATLAS_XV_Michele_2013} and  2\% on the Einstein radius \citep[e.g.,][]{Bolton08}.  We neglected possible contributions to the lensing quantities by external convergence because it is at a few percent level at most \citep{birrer20} and would not affect our conclusions in a significant way.

In addition to the scaling relations described by \citet{Auger_2010}, we considered a new correlation that involves the mass-density profile slope $\gamma_{\rm lens}$ determined from lens models \citep[]{T24, William_2025_dinos_2}. This quantity was not available to \citet{Auger_2010}. We note that \citet{Auger_2010} used a slope $\gamma$ derived by a joint lensing and dynamical analysis, which is the same only when the mass-density profiles are exactly power laws \citep{Sonnenfeld13b}. All the quantities presented in this analysis are summarized in Table~\ref{tab:data}.

\subsection{Correlation between the lensing-mass profile slope and ratio of the lensing and stellar velocity dispersion}

The first correlation that we examined is the correlation between the lensing slope and the ratio of the stellar velocity dispersion $\sigma$ and the lensing velocity dispersion $\sigma_{\rm SIE}$, $f_{\rm SIE}=\sigma/\sigma_{\rm SIE}$. The lensing velocity dispersion is related to the Einstein radius $\theta_{\rm E}$ and the angular diameter distance to the source (D$_{\rm s}$) and between the deflector and the source (D$_{\rm ds}$) by the equation \citep{Treu10b}
\begin{equation}
\theta_E=4\pi\left(\frac{\sigma_{\rm
SIE}}{c}\right)^2\frac{D_{ds}}{D_s}.
\label{eq:sSIE}
\end{equation}
The subscript SIE is used as a reminder that the Einstein radius is usually derived by fitting a singular isothermal ellipsoid to the lensing features. It is worth noting, however, that the Einstein radius is very robust to the choice of lens model and only varies by 1-2\% when a power-law model is adopted instead \citep{Bolton08b}.

\citet{Treu06} introduced $\fsie$ as a proxy for the slope of the mass density profile, finding that it should be close to unity, as expected for an isothermal mass density profile. This can be understood qualitatively in terms of spherical profiles. Given a power-law total mass-density profile with the three-dimensional form $\rho_{\rm tot} \propto r^{-\gamma}$, the mass enclosed within a two-dimensional circular aperture with radius $R$ is given by 
\begin{equation}
M (<R) = M_{\rm E} \left(\frac{R}{R_{\rm E}}\right)^{3-\gamma}, 
\label{eq:2dmass}
\end{equation}
where R$_{\rm E}$ is the Einstein radius in kpc, and M$_{\rm E}$ is the mass within the Einstein radius. Following \citet{Schneider06}, it is easy to show that 
\begin{equation}
\sigma_{\rm SIE}^{2} R_{\rm E} = \frac{G M_{\rm E}}{\pi}.
\label{eq:vSIE}
\end{equation}
The virial theorem holds that 
\begin{equation}
\sigma_{1/2}^{2} R_{\rm eff}/2 = \frac{G M(<R_{\rm eff}/2)}{c_{\rm e2}},
\label{eq:ce2}
\end{equation}
where R$_{\rm eff}$ is the effective radius, and $c_{\rm e2}$ is a dimensionless structural parameter that depends on the three-dimensional shape of the mass distribution, orbital structure, and spatial distribution of luminous tracers \citep{NTB08}. \citet{Bolton08b} finds $\langle \log c_{\rm e2}\rangle=0.53\pm0.06$. In the following, we assume that c$_{\rm e2}$ is a constant, which is a good approximation.
Residual trends from the predicted relations would imply that c$_{\rm e2}$ is not exactly a constant, but that it depends on galaxy mass, stellar velocity dispersion, and the mass-density profile slope, for instance.

By combining Eqs. (~\ref{eq:2dmass}) to (~\ref{eq:ce2}), we find that
\begin{equation}
f_{\rm SIE}^2 = \frac{\pi}{c_{\rm e2}}\left(\frac{R_{\rm eff}/2}{R_{\rm E}}\right)^{2-\gamma}.
\label{eq:fsie2}
\end{equation}

\begin{figure}
\includegraphics[width=\columnwidth]{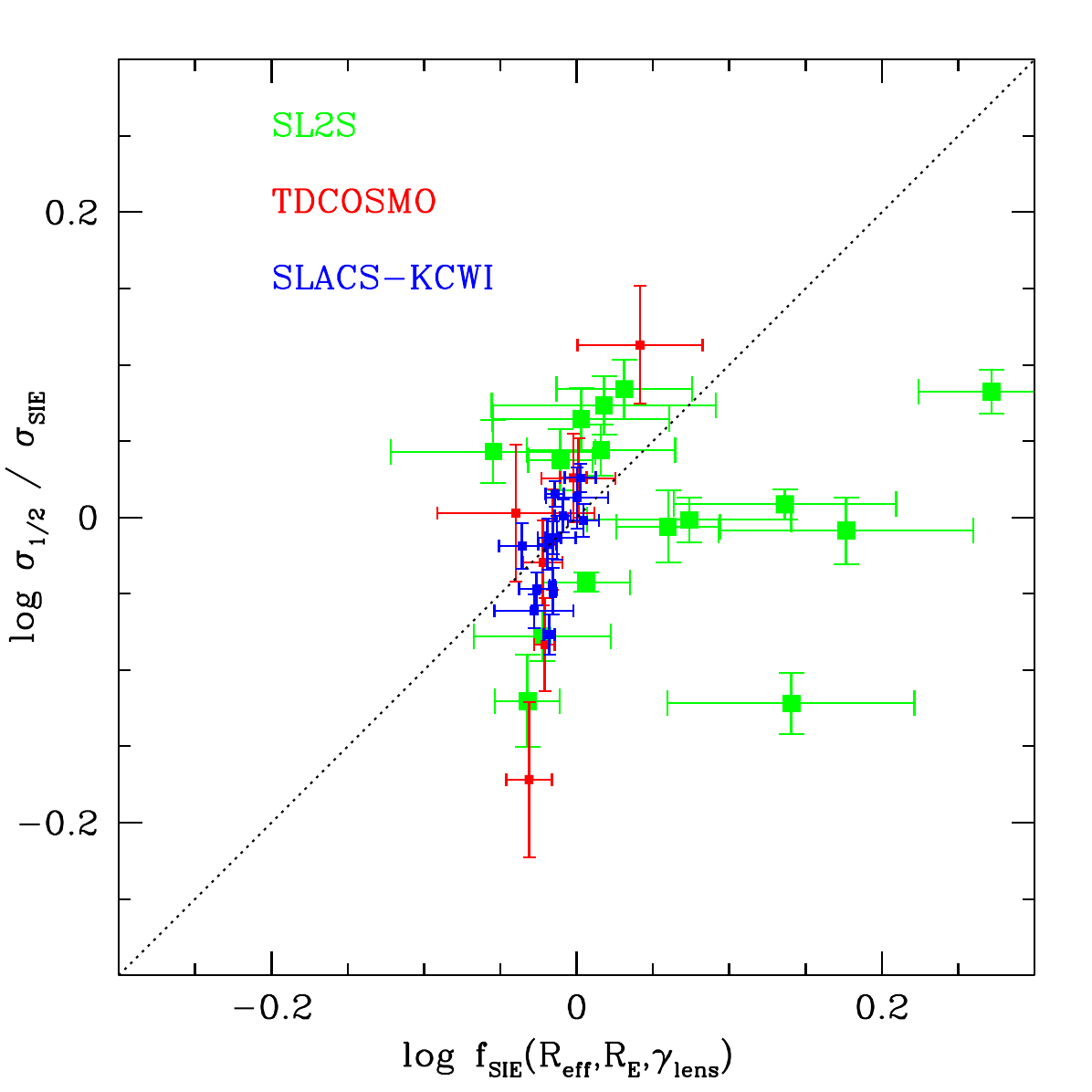}
	\caption{\label{fig:fsie} Comparison between the measured $f_{\rm SIE}$ and that derived from Eq. (~\ref{eq:fsie2}).}
\end{figure}

In Fig.~\ref{fig:fsie} we compare the measured f$_{\rm SIE}$ with the values predicted by Eq. (~\ref{eq:fsie2}), using c$_{\rm e2}$ measured by \citet{NTB08}. Even though the data are noisy, we note that the SLACS-KCWI and TDCOSMO samples are very close to the predicted values. There is more scatter for the SL2S, which is in part expected because $R_{\rm eff} \sim R_{\rm E} /2$, and thus, the factor to be elevated to the power of $2-\gamma_{\rm lens}$ is 4 on average, introducing extrapolation errors. Furthermore, c$_{e2}$ is not a constant and is expected to depend significantly on $\gamma$. Thus, using the average value is accurate for systems with $\gamma\sim2$, similar to the SLACS sample and less so when $\gamma$ departs from 2.

\subsection{The fundamental planes of the lensing mass}

\citet{Bolton07} introduced the lensing-mass fundamental plane \citep[MFP; see also][]{Bolton08b,Auger_2010} by analogy with the standard fundamental plane, that is, they replaced the surface brightness with the surface-mass density,
\begin{equation}
\log R_{\rm eff} = a \log \sigma_{1/2} + b \log \Sigma_{1/2} + c. 
\label{eq:MFP}
\end{equation}

This MFP could provide some useful insight. For a power law, the surface-mass density within a projected radius is given by
\begin{equation}
\Sigma=\Sigma_c \left(\frac{R}{R_{\rm E}}\right)^{1-\gamma},
\end{equation}
where $\Sigma_c$ is the critical surface mass density
\begin{equation}
\Sigma_c=\frac{M_{\rm E}}{\pi R_{\rm E}^2}=\frac{c^2 D_s}{4\pi G D_d D_{ds}}.
\end{equation}

Thus, Eq. (~\ref{eq:ce2}) can be rewritten as 
\begin{equation}
\sigma_{1/2}^2 = \frac{\pi G \Sigma_{1/2} R_{\rm eff}}{2 c_{\rm e2}},
\end{equation}
where $\Sigma_{1/2}$ is the average surface-mass density within half the effective radius.
Taking the logarithm of both sides and sorting the variables yields
\begin{equation}
\log R_{\rm eff} = 2 \log \sigma_{1/2} - \log \Sigma_{1/2} - \log \frac{\pi G}{2 c_{\rm e2}}.
\label{eq:mpft}
\end{equation}
The coefficients are found to be close to $a=2$ and $b=-1$, indicating that the virial coefficient is close to a constant, and thus, the lensing-mass fundamental plane  is much better described by this simple interpretation than by the classic fundamental plane, which appears to be tilted with respect to the virial prediction. This requires variations in the M/L ratio with $\sigma$, which are generally attributed to changes in the dark matter content, stellar population age, metallicity, and initial mass function \citep{CLR96,TBB04,Dutton13,Bernardi20,Deugenio21,Treu06,Treu10,Auger10}. The intercept depends on the units. Consistent results, with coefficients in agreement with virial predictions, are obtained when the mass plane (MP) is measured using mass from stellar dynamics along with integral-field stellar kinematics \citep[e.g.,][]{Cappellari2006, ATLAS_XV_Michele_2013,Zhu2024}.

\begin{figure}
\includegraphics[width=\columnwidth]{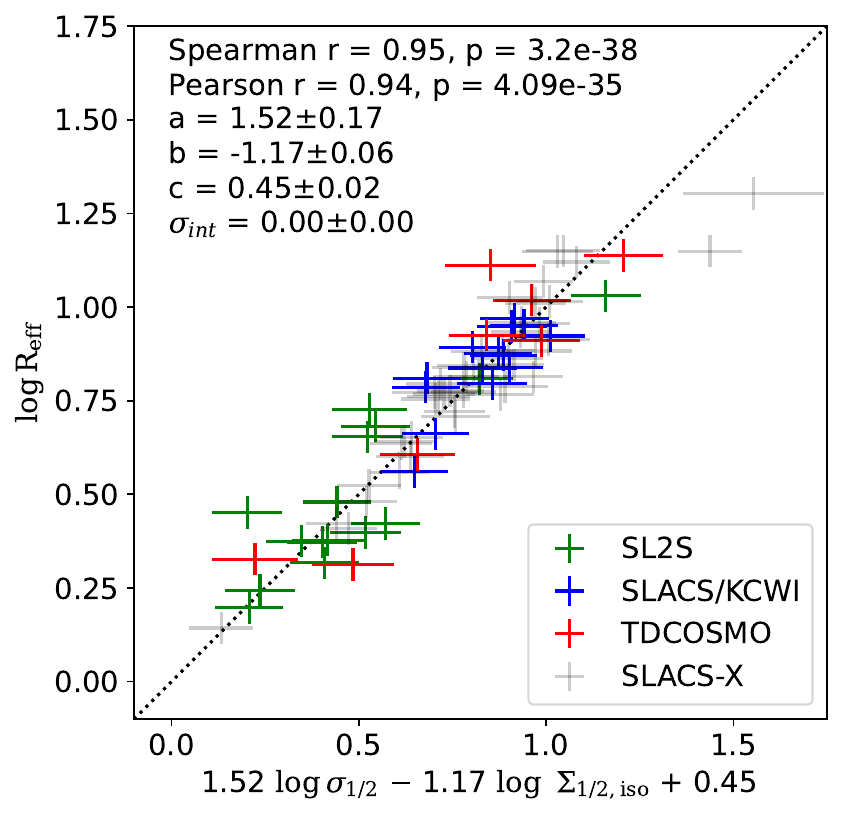}
	\caption{Lensing-mass fundamental plane with isothermal mass profile assumption, $\mathrm{ \log R_{eff} = a \log \sigma_{1/2} + b \log \Sigma_{1/2,iso} + c}$. The plane is fit to the combined SL2S, SLACS/KCWI, and SLACS-X samples. The TDCOSMO sample is overplotted. Here, $\mathrm{\Sigma_{1/2,iso}=M_{R_{eff}/2, iso}/\pi(R_{eff}/2)^2}$ is the average lensing surface-mass density within half the effective radius, assuming the mass density profile is isothermal (i.e., $\gamma_{lens} = 2$). $\sigma_{1/2}$ is given in units of 250 $\mathrm{ km \ s^{-1} }$, $\mathrm{M_{1/2}}$ in units of $10^{10} \ \mathrm{M_{\odot}}$, and $\mathrm{R_{eff}}$ in kpc. The dotted line shows the one-to-one line. \label{fig:isomfp}
	}
\end{figure}

Previous studies \citep{Bolton07,Bolton08,Auger_2010} computed the lensing-mass fundamental plane assuming isothermal mass profiles ($\gamma=2$). In Fig.~\ref{fig:isomfp} we derive the MFP under the same assumption, using the SL2S sample, SLACS-KCWI sample, and the SLACS sample reported by \citet{Auger_2010} based on SDSS velocity dispersions (SLACS-X). To fit the plane, we used the Python package {\sc LtsFit } \citep{ATLAS_XV_Michele_2013}, a robust least-squares regression with uncertainties in all dimensions and intrinsic scatter. We did not clip outliers, but included all data points in the fit. Effective radii were fit in kpc, velocity dispersions in units of 250 $\mathrm{km \ s^{-1}}$, and surface-mass densities in $\mathrm{10^{10} \ M_{\odot} \ kpc^{-2}}$. The samples followed the same MFP with coefficients consistent with those found by \citet{Auger_2010}. The slopes are close but not identical to those in Eq. (~\ref{eq:mpft}), indicating that the assumption of a constant virial coefficient is not exactly correct, but it is likely to have some dependence on mass or velocity dispersion because elliptical galaxies are not completely scale invariant. We overplot the TDCOSMO sample for illustration and show that it lies along the plane across the sample space occupied by the SL2S and SLACS samples. The fit indicates no intrinsic scatter to the combined sample. As a test, we fitted all combinations of subsamples, and the coefficients are the same within the errors.

We also studied for the first time the MFP for a generic power-law profile using the measurements of $\gamma=\gamma_{\rm lens}$ that are now available. The uncertainty on $\Sigma_{1/2}$ is given by
\begin{equation}
\delta \log \Sigma_{1/2} = \sqrt{(1-\gamma_{\rm lens})^2\left(\frac{\delta R_{\rm eff}}{R_{\rm eff} \ln 10}\right)^2+\delta \gamma^2 \left(\log \frac{R_{\rm eff}}{2 R_E}\right)^2} 
\label{eq:dsigma}.
\end{equation}

\begin{figure}
\includegraphics[width=\columnwidth]{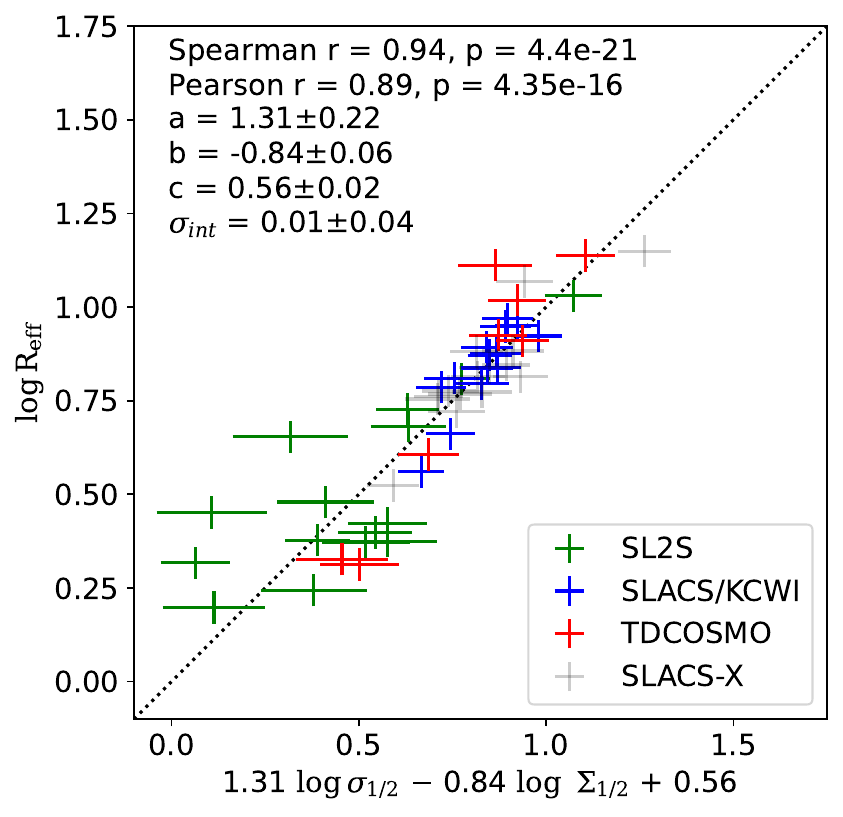}
\caption{Lensing-mass fundamental plane with a flexible power-law mass profile $\mathrm{ \log R_{eff} = a \log \sigma_{1/2} + b \log \Sigma_{1/2} + c}$. The plane is fit to the combined SL2S, SLACS/KCWI, and SLACS-X samples. The TDCOSMO sample is overplotted.
Here $\mathrm{\Sigma_{1/2}=M_{R_{eff}/2}/\pi(R_{eff}/2)^2}$ is the average lensing surface-mass density within half the effective radius. We updated the SLACS-X lens models to those from \citet{T24} with free power-law slopes to compare with SL2S and SLACS/KCWI lens models. $\sigma_{1/2}$ is given in units of 250 $\mathrm{ km \ s^{-1} }$, $\mathrm{M_{1/2}}$ in units of $10^{10} \ \mathrm{M_{\odot}}$, and $\mathrm{R_{eff}}$ in kpc. The dotted line shows the one-to-one line.\label{fig:gammamfp}
}
\end{figure}

{The resulting MFP is shown in Fig.~\ref{fig:gammamfp}.} As expected, the errors are larger than in the isothermal version, especially for the SL2S sample. For this sample,  $\delta \gamma_{\rm lens}$ is generally larger than for TDCOSMO and SLACS because the S/N of the imaging data is lower. Furthermore, the Einstein radius is typically several times the effective radius, requiring more extrapolation than for the other samples. This plane is consistent with no scatter within its error bars as well, but in Fig.~\ref{fig:gammamfp} we report the best fit, which includes an intrinsic scatter of 1\%. We also examined the scaling relations using the previously measured velocity dispersions for the SL2S subsample while retaining the other sample as before. The results were nearly identical: all coefficients agreed to within half a sigma. When the SL2S sample was excluded, the fit to the SLACS samples indicated no intrinsic scatter, and the coefficients agreed to within 1$\sigma$. Of the 74 galaxies we fit to evaluate the scaling relations, only 15 are from the SL2S sample. We therefore expect its impact to be negligible.

The two planes are both well defined, and there are no significant differences between the subsamples. We quantified any residual differences between the subsamples in the following way. Since the subsamples, except SLACS-X, are too small to robustly determine the slopes, we kept the slopes $a$ and $b$ fixed to the values obtained for the combined sample, and we then fit for the intercept $c$ of each subsample. In the isothermal case, the intercept is $0.451\pm0.019$, $0.463\pm0.027$, $0.456\pm0.027$, and $0.445\pm0.015$ for the combined, SL2S, SLACS-KCWI, and SLACS-X samples, respectively. They are consistent within the errors. The error on the intercept is largest for the SL2S sample and corresponds to $\pm$5.8\% in R$_{\rm eff}$ and $\pm4.2\%$ in $\sigma_{1/2}$. For the nonisothermal case, the intercept is $0.560\pm0.020$, $0.579\pm0.029$, $0.561\pm0.021$, and $0.548\pm0.022$ for the samples in the same order as above. Again, they are consistent within the errors. The error is largest for the SL2S sample and corresponds to $\pm$6.9\% in R$_{\rm eff}$ and $\pm5.3\%$ in $\sigma_{1/2}$.

\begin{figure}
\includegraphics[width=\columnwidth]{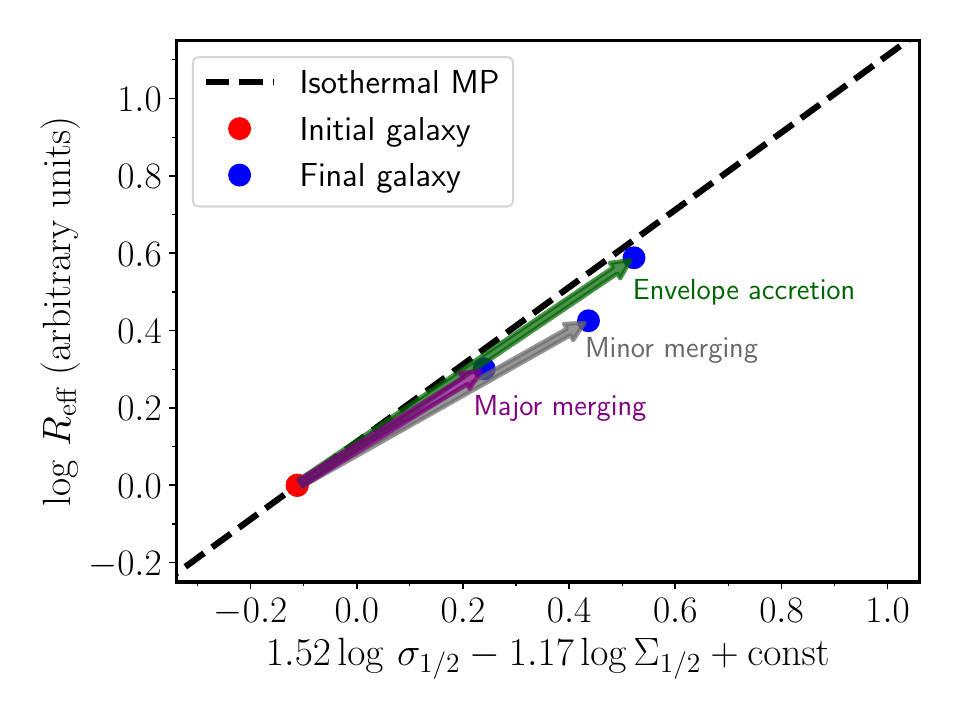}
	\caption{Evolution in the same parameter space as in Fig.~\ref{fig:isomfp} of a reference galaxy that grows by a factor of two in stellar mass according to three different processes: Envelope accretion (green arrow), and major (purple arrow) and minor (gray arrow) dry merging.  In all cases, the initial galaxy (red symbol) is assumed to lie on the isothermal MP (dashed line). The final position (blue symbol) of the galaxy in the diagram was calculated using the model described in section 6.2 of \citet{Nipoti2025} for envelope accretion and using the equations of section 3.1 of \citet{Nipoti2025} for minor (mass ratio $\xi=0.1$) and major (mass ratio $\xi=1$) dry merging, assuming a size-mass relation slope $\beta_R=0.76$ (see \citealt{Nipoti2025}).  \label{fig:isomf-theory}
	}
\end{figure}

In conclusion, the MFP analysis shows no evidence of structural differences between the SL2S, SLACS, and TDCOSMO samples. This does not mean that there is no evolution in the characteristics of the galaxies. For example, in Figs.~\ref{fig:isomfp} and~\ref{fig:gammamfp}, the SL2S galaxies tend to have smaller effective radii, which is generally consistent with the finding that early-type galaxies have smaller sizes at high redshift than in the local Universe for a given stellar mass \citep[e.g.,][]{Vd10}. \citet{Sonnenfeld15} found that the fraction of dark matter within the inner 5 kpc increases with redshift, and \citet{William_2025_dinos_2} also found mild evidence for evolution. We showed, however, that galaxies evolve within the lensing-mass fundamental plane. An example of a mechanism that moves galaxies along the lensing-mass plane is the so-called envelope accretion that was recently considered by \citet{Nipoti2025} in addition to the standard major and minor mergers. Envelope accretion describes the case in which diffuse satellites are completely disrupted in the galaxy outskirts. 
As shown in Fig.~\ref{fig:isomf-theory}, the model proposed by \citet{Nipoti2025} tantalizingly predicts that envelope accretion causes galaxies to evolve almost exactly along the lensing-mass plane. Similar results were obtained for the generalized mass plane with free mass-density slopes.
For comparison, we show in the same figure the predicted evolution of galaxies that grow in mass by a factor of two via either minor merging with a mass ratio 1/10 or by equal-mass major dry merging, as predicted by standard toy models (e.g., \citealt{Nipoti2025}). The galaxy follows paths that lie almost parallel to the isothermal MP in these cases as well. However, they form slightly larger angles with the isothermal MP than the angle that formed by the path predicted by the envelope-accretion model.
A full comparison with galaxy evolution models is beyond the scope of this paper and is left for future work.

\section{Summary}
\label{sec:summary}

We have applied improved methods and clean stellar template libraries \citep{Knabel_Mozumdar_2025} to obtain new stellar velocity dispersions using spectra that were obtained at the Keck and VLT Telescopes for the lens galaxies in the SL2S sample \citep{Gavazzi12,Ruff11,Sonnenfeld_2013a,Sonnenfeld13b,Sonnenfeld15, William_2025_dinos_2}. The SL2S lens sample matches the quasar lenses used by the TDCOSMO collaboration for time-delay cosmography very well in terms of deflector redshift and Einstein radii   \citep{Millon_2020_TDCOSMO_I}, and it can be used as an external dataset \citep{birrer20} to improve the constraints on the Hubble constant and other cosmological parameters \citep{B+T21}, provided stellar velocity dispersions are accurately measured.  We used these new measurements to investigate the lensing-mass fundamental plane \citep{Bolton07,Bolton08,Auger_2010} as a tool to compare the internal structure of massive galaxies. Our main results are summarized below.

\begin{itemize}
\item The new measurements are consistent on average with the old measurements, but there is significant scatter for individual systems.  
\item Systematic errors are dominated by the choice of stellar templates and are about 2-3\%, depending on the S/N of the spectra.
\item Stellar templates introduce a positive covariance between velocity dispersion measurements in the galaxy samples. The covariance is at the level of 1-2\%, depending on spectral quality, for galaxies that were observed with the same instrumental setup. It is significantly smaller (sub-percent) for spectra that were obtained with different instrumental setups.
\item We derived a simple formula (Eqs. \ref{eq:mpft} and \ref{eq:dsigma}) that relates the slope of the mass-density profile to the observed stellar velocity dispersion, Einstein radius, and effective radius under the assumption of a constant virial coefficient, and we showed that it is consistent within the errors with the observations.
\item The SL2S, SLACS, and TDCOSMO galaxies follow a tight fundamental plane of the lens mass that is consistent within the errors with the same plane derived for the SLACS galaxies alone.
\item We introduced a new version of the lensing-mass fundamental plane by relaxing the assumption of isothermal mass-density profiles. This new version is also well defined, but the error bars are larger, as expected, and the three samples are mutually consistent within $1\sigma$ uncertainty.
\end{itemize}

In terms of the larger picture, the two main conclusions of this study are that first, with the new kinematic measurements and systematic error estimates, the SL2S galaxies with the best spectra can be used as external datasets for time-delay cosmography \citep[as discussed by][]{Milestone25}. Second, the internal mass structure of massive elliptical galaxies evolves with redshift. This evolution occurs within the lensing-mass fundamental plane, however, which sets tight constraints on their evolutionary mechanism, such as dry mergers \citep{NTB09} or envelope accretion \citep{Nipoti2025}. 

\begin{acknowledgements}
The first two authors should be regarded as joint first authors. We are grateful to William Sheu and our friends and colleagues of the TDCOSMO collaboration for many useful discussions on these topics.
PM, SK, and TT acknowledge support by NSF through grants NSF-AST-1906976, NSF-AST-1836016, and NSF-AST-2407277, and from the Moore Foundation through grant 8548.
AJS received support from NASA through the STScI grants HST-GO-16773 and JWST-GO-2974.
\end{acknowledgements}

\bibliographystyle{aa} 
\bibliography{main} 

\onecolumn

\begin{appendix}
\section{Measured quantities for SL2S and data from external samples}
\renewcommand{\arraystretch}{1.2}
\setlength{\tabcolsep}{4pt}
\begin{longtable}{l c c c c c c c c c}
\multicolumn{10}{c}{\parbox{\textwidth}{\textbf{Table \ref{table:veldis_measurements}.} Measured values for the SL2S sample.}}\\
\label{table:veldis_measurements}\\

\hline \hline
Name  &   Instrument &  S/N & $\sigma_{\text{inst}}$ & $\overline{\sigma}_{\text{BIC\_w}}$ &  $\delta \overline{\sigma}_{\text{BIC\_w}}$ & $\overline{\sigma}_{\text{eq\_w}}$ & $\delta \overline{\sigma}_{\text{eq\_w}}$ &  $\Delta_B \overline{\sigma}_{\text{BIC\_w, eq\_w}}$  &  $\frac{\delta \sigma_{\text{inst}}}{\sigma_{\text{inst}}} (\frac{\sigma_{\text{inst}}}{\sigma})^2$ \\
   &  &  (\AA$^{-1}$) & (\kmps) & (\kmps) &  (\kmps) &  (\kmps)  &  (\kmps)  &  (\kmps)  &   \\
\hline
\endfirsthead

\multicolumn{10}{c}{\textbf{Table \ref{table:veldis_measurements} continued.}} \\
\hline \hline
Name    &   Instrument &  S/N & $\sigma_{\text{inst}}$ & $\overline{\sigma}_{\text{BIC\_w}}$ &  $\delta \overline{\sigma}_{\text{BIC\_w}}$ & $\overline{\sigma}_{\text{eq\_w}}$ & $\delta \overline{\sigma}_{\text{eq\_w}}$ &  $\Delta_B \overline{\sigma}_{\text{BIC\_w, eq\_w}}$  &  $\frac{\delta \sigma_{\text{inst}}}{\sigma_{\text{inst}}} (\frac{\sigma_{\text{inst}}}{\sigma})^2$ \\
   &  &  (\AA$^{-1}$) & (\kmps) & (\kmps) &  (\kmps) &  (\kmps)  &  (\kmps)  &  (\kmps)  &   \\
\hline
\endhead

\hline
\endfoot

\hline
\multicolumn{10}{c}{}{\parbox{\textwidth}{\vspace{2pt}\textbf{Notes.} For each lens we list the instruments used, S/N per \AA\ of the spectra, measured instrumental resolution, BIC weighted and equally weighted velocity dispersions, and associated statistical and systematic uncertainties. The BIC determined weights used for each galaxy are reported in Table~\ref{table:results}. The systematic uncertainties are Bessel corrected. The last column shows the relative error associated with uncertainties in the measurement of instrumental resolution. For galaxies with two spectra of sufficient spectral resolution, we report both values. The one with the highest S/N is listed first, and it is our fiducial measurement. }}
\endlastfoot
SL2SJ020833-071414 & LRIS & 14 & 64 $\pm$ 6 & 252 &  19 & 252 & 19 & 3  &  0.006 \\
SL2SJ021206-075528 & LRIS & 23 & 52 $\pm$ 7 & 236 &  7 & 235 & 8 & 2  &  0.006 \\
SL2SJ021247-055552 & XSHOOTER & 15 & 32 $\pm$ 2 & 256 &  10 & 254 & 10 & 4  &  0.001 \\
SL2SJ021325-074355 & LRIS & 4 & 97 $\pm$ 8 & 344 &  24 & 352 & 24 & 13  &  0.007 \\
SL2SJ021411-040502 & XSHOOTER & 23 & 32 $\pm$ 2 & 209 &  7 & 213 & 7 & 7  &  0.001 \\
                   & LRIS & 10 & 59 $\pm$ 6 & 250 &  15 & 254 & 15 & 7  &  0.006 \\
SL2SJ021737-051329 & LRIS & 9 & 97 $\pm$ 8 & 253 &  20 & 258 & 19 & 8  &  0.012 \\
                   & LRIS & 8 & 68 $\pm$ 7 & 248 &  14 & 249 & 14 & 2  &  0.008 \\
SL2SJ021801-080247 & XSHOOTER & 5 & 31 $\pm$ 2 & 254 &  28 & 255 & 28 & 3  &  0.001 \\
SL2SJ021902-082934 & LRIS & 17 & 99 $\pm$ 4 & 285 &  7 & 284 & 7 & 7  &  0.005 \\
SL2SJ022046-094927 & XSHOOTER & 8 & 32 $\pm$ 2 & 187 &  16 & 188 & 16 & 1  &  0.002 \\
SL2SJ022056-063934 & LRIS & 18 & 97 $\pm$ 8 & 246 &  11 & 245 & 11 & 8  &  0.013 \\
SL2SJ022346-053418 & LRIS & 15 & 64 $\pm$ 6 & 298 &  16 & 299 & 16 & 7  &  0.004 \\
SL2SJ022357-065142 & LRIS & 20 & 59 $\pm$ 4 & 254 &  16 & 253 & 16 & 4  &  0.003 \\
                   & LRIS & 18 & 68 $\pm$ 7 & 258 &  20 & 270 & 21 & 19  &  0.007 \\
SL2SJ022610-042011 & LRIS & 20 & 94 $\pm$ 6 & 310 &  13 & 311 & 13 & 6  &  0.006 \\
                   & LRIS & 13 & 93 $\pm$ 9 & 287 &  11 & 284 & 11 & 5  &  0.01 \\
SL2SJ022648-040610 & XSHOOTER & 31 & 32 $\pm$ 2 & 337 &  9 & 343 & 8 & 8  &  0.001 \\
SL2SJ022648-090421 & LRIS & 16 & 97 $\pm$ 8 & 326 &  8 & 323 & 8 & 12  &  0.007 \\
                   & LRIS & 7 & 68 $\pm$ 7 & 350 &  24 & 353 & 23 & 5  &  0.004 \\
SL2SJ023251-040823 & XSHOOTER & 35 & 30 $\pm$ 1 & 208 &  5 & 208 & 5 & 1  &  0.001 \\
                   & LRIS & 16 & 93 $\pm$ 9 & 240 &  9 & 237 & 9 & 4  &  0.015 \\
SL2SJ084847-035103 & XSHOOTER & 13 & 33 $\pm$ 2 & 198 &  8 & 198 & 8 & 2  &  0.002 \\
SL2SJ084909-041226 & DEIMOS & 17 & 61 $\pm$ 6 & 338 &  14 & 340 & 14 & 5  &  0.003 \\
SL2SJ084934-043352 & LRIS & 19 & 56 $\pm$ 2 & 221 &  11 & 221 & 11 & 0  &  0.003 \\
SL2SJ084959-025142 & XSHOOTER & 65 & 33 $\pm$ 0 & 208 &  4 & 208 & 4 & 1  &  0.0 \\
SL2SJ085019-034710 & LRIS & 20 & 58 $\pm$ 1 & 234 &  12 & 232 & 12 & 10  &  0.001 \\
SL2SJ085540-014730 & LRIS & 18 & 56 $\pm$ 2 & 196 &  11 & 195 & 10 & 1  &  0.003 \\
SL2SJ085559-040917 & LRIS & 27 & 51 $\pm$ 7 & 275 &  8 & 279 & 8 & 11  &  0.005 \\
SL2SJ085826-014300 & LRIS & 8 & 64 $\pm$ 6 & 240 &  14 & 241 & 14 & 2  &  0.007 \\
SL2SJ090106-025906 & XSHOOTER & 5 & 33 $\pm$ 2 & 224 &  25 & 219 & 25 & 27  &  0.001 \\
SL2SJ090407-005952 & XSHOOTER & 19 & 33 $\pm$ 1 & 232 &  9 & 233 & 9 & 0  &  0.001 \\
SL2SJ095921+020638 & XSHOOTER & 17 & 33 $\pm$ 1 & 200 &  8 & 199 & 8 & 0  &  0.001 \\
SL2SJ135847+545913 & LRIS & 22 & 62 $\pm$ 9 & 256 &  9 & 256 & 9 & 1  &  0.008 \\
SL2SJ135949+553550 & LRIS & 9 & 65 $\pm$ 6 & 221 &  14 & 222 & 15 & 1  &  0.008 \\
                   & LRIS & 7 & 68 $\pm$ 7 & 219 &  20 & 219 & 19 & 0  &  0.01 \\
SL2SJ140123+555705 & LRIS & 9 & 85 $\pm$ 13 & 379 &  18 & 371 & 18 & 13  &  0.008 \\
SL2SJ140156+554446 & LRIS & 28 & 64 $\pm$ 5 & 276 &  9 & 275 & 9 & 3  &  0.004 \\
SL2SJ140454+520024 & LRIS & 31 & 63 $\pm$ 6 & 336 &  9 & 342 & 9 & 15  &  0.003 \\
SL2SJ140546+524311 & LRIS & 19 & 61 $\pm$ 8 & 258 &  8 & 258 & 8 & 2  &  0.007 \\
SL2SJ140614+520253 & LRIS & 8 & 85 $\pm$ 13 & 268 &  14 & 272 & 14 & 6  &  0.015 \\
SL2SJ140650+522619 & LRIS & 12 & 65 $\pm$ 6 & 270 &  11 & 267 & 11 & 6  &  0.005 \\
SL2SJ142031+525822 & LRIS & 19 & 70 $\pm$ 3 & 218 &  11 & 216 & 11 & 7  &  0.005 \\
SL2SJ142059+563007 & LRIS & 20 & 63 $\pm$ 8 & 244 &  16 & 247 & 16 & 7  &  0.008 \\
                   & LRIS & 15 & 68 $\pm$ 7 & 250 &  10 & 250 & 10 & 0  &  0.008 \\
SL2SJ220329+020518 & LRIS & 29 & 66 $\pm$ 2 & 203 &  10 & 208 & 10 & 10  &  0.002 \\
SL2SJ220506+014703 & XSHOOTER & 31 & 32 $\pm$ 2 & 319 &  7 & 309 & 7 & 14  &  0.001 \\
SL2SJ220629+005728 & LRIS & 4 & 97 $\pm$ 8 & 281 &  27 & 286 & 27 & 9  &  0.01 \\
SL2SJ221326-000946 & XSHOOTER & 30 & 31 $\pm$ 2 & 194 &  12 & 193 & 11 & 1  &  0.002 \\
                   & LRIS & 23 & 72 $\pm$ 7 & 162 &  14 & 165 & 13 & 5  &  0.019 \\
SL2SJ221407-180712 & LRIS & 5 & 97 $\pm$ 8 & 200 &  16 & 208 & 16 & 13  &  0.019 \\
SL2SJ221852+014038 & LRIS & 19 & 64 $\pm$ 3 & 292 &  11 & 291 & 11 & 4  &  0.002 \\
SL2SJ221929-001743 & LRIS & 26 & 55 $\pm$ 3 & 242 &  6 & 241 & 7 & 0  &  0.002 \\
                   & LRIS & 32 & 93 $\pm$ 9 & 239 &  11 & 242 & 11 & 6  &  0.015 \\
SL2SJ222012+010606 & DEIMOS & 17 & 79 $\pm$ 1 & 248 &  15 & 248 & 14 & 1  &  0.002 \\
SL2SJ222148+011542 & DEIMOS & 29 & 74 $\pm$ 5 & 277 &  10 & 283 & 10 & 8  &  0.005 \\
SL2SJ222217+001202 & LRIS & 11 & 63 $\pm$ 4 & 198 &  20 & 196 & 20 & 3  &  0.006 \\
                   & LRIS & 8 & 68 $\pm$ 7 & 190 &  28 & 195 & 29 & 8  &  0.013 \\
\end{longtable}
\twocolumn

\renewcommand{\arraystretch}{1.15}
\begin{table*}
\caption{Values used in scaling relation calculations and figures.}
\begin{center}
\begin{threeparttable}
 \resizebox{0.85\textwidth}{!}{
\begin{tabular}{l c c c c c c c} 
    \hline
Name & $\mathrm{ z_{d} }$ & $\mathrm{ z_{s} }$ & $\mathrm{ 	R_{eff} }$ & $\mathrm{ 	\theta_E }$ & $\mathrm{ \sigma_{1/2} }$ & $\mathrm{ \gamma_{lens} }$ \\
& & & $\left( \arcsec \right)$ & $\left( \arcsec \right)$ & $\mathrm{ \left(km \ s^{-1} \right)}$ & & \\
\hline
\midrule
 SL2SJ021411-040502 & 0.609 & 1.88 & 0.42 & 1.34  $\pm$ 0.03 & 219  $\pm$ 10 & 2.39  $\pm$ 0.2 \\
 SL2SJ021902-082934 & 0.389 & 2.15 & 0.39 & 1.30  $\pm$ 0.03 & 303  $\pm$ 10 & 2.70  $\pm$ 0.11 \\
 SL2SJ022610-042011 & 0.494 & 1.23 & 0.39 & 1.15  $\pm$ 0.02 & 324  $\pm$ 15 & 2.05  $\pm$ 0.15 \\
 SL2SJ023251-040823 & 0.352 & 2.34 & 0.32 & 1.02  $\pm$ 0.02 & 221  $\pm$ 5 & 2.38  $\pm$ 0.18 \\
 SL2SJ084909-041226 & 0.722 & 1.54 & 0.24 & 1.14  $\pm$ 0.03 & 363  $\pm$ 16 & 2.07  $\pm$ 0.15 \\
 SL2SJ084959-025142 & 0.274 & 2.09 & 1.54 & 1.19  $\pm$ 0.01 & 207  $\pm$ 3 & 2.24  $\pm$ 0.30 \\
 SL2SJ085540-014730 & 0.365 & 3.39 & 0.47 & 0.96  $\pm$ 0.02 & 203  $\pm$ 11 & 2.25  $\pm$ 0.11 \\
 SL2SJ090407-005952 & 0.611 & 2.36 & 0.71 & 1.41  $\pm$ 0.03 & 238  $\pm$ 9 & 1.98  $\pm$ 0.15 \\
 SL2SJ095921+020638 & 0.552 & 3.35 & 0.41 & 0.71  $\pm$ 0.02 & 209  $\pm$ 8 & 2.12 $\pm$ 0.18 \\
 SL2SJ140454+520024 & 0.456 & 1.59 & 0.78 & 2.39  $\pm$ 0.05 & 360  $\pm$ 18 & 2.49  $\pm$ 0.21 \\
 SL2SJ140546+524311 & 0.526 & 3.01 & 0.48 & 1.46  $\pm$ 0.03 & 270  $\pm$ 9 & 2.23  $\pm$ 0.17 \\
 SL2SJ142059+563007 & 0.483 & 3.12 & 1.79 & 1.39  $\pm$ 0.01 & 283  $\pm$ 13 & 2.06  $\pm$ 0.22 \\
 SL2SJ220329+020518 & 0.400 & 2.15 & 0.99 & 1.68  $\pm$ 0.03 & 217  $\pm$ 15 & 1.94  $\pm$ 0.08 \\
 SL2SJ220506+014703 & 0.476 & 2.53 & 0.40 & 1.81  $\pm$ 0.04 & 333  $\pm$ 16 & 1.92  $\pm$ 0.14 \\
 SL2SJ222148+011542 & 0.325 & 2.35 & 0.53 & 1.27  $\pm$ 0.03 & 290  $\pm$ 13 & 2.14  $\pm$ 0.13 \\
\hline
 SDSSJ002907.77-005550.5 & 0.227 & 0.931 & 2.30 & 0.95  $\pm$ 0.02 & 207  $\pm$ 7 & 2.47  $\pm$ 0.26 &  \\
 SDSSJ003753.21-094220.1 & 0.195 & 0.632 & 2.30 & 1.48  $\pm$ 0.03 & 279  $\pm$ 7 & 2.38  $\pm$ 0.11 &   \\
 SDSSJ033012.14-002051.9 & 0.351 & 1.071 & 1.26 & 1.09  $\pm$ 0.02 & 258  $\pm$ 12 & 2.14  $\pm$ 0.17 &   \\
 SDSSJ111250.60+082610.4 & 0.273 & 0.629 & 1.55 & 1.50  $\pm$ 0.03 & 274  $\pm$ 7 & 1.92  $\pm$ 0.18 &   \\ SDSSJ120444.07+035806.4 & 0.164 & 0.631 & 1.63 & 1.30  $\pm$ 0.03 & 262  $\pm$ 5 & 2.02  $\pm$ 0.06 &   \\
 SDSSJ125028.26+052349.1 & 0.232 & 0.795 & 1.86 & 1.12  $\pm$ 0.02 & 243  $\pm$ 6 & 2.18  $\pm$ 0.09 &   \\
 SDSSJ130613.65+060022.1 & 0.173 & 0.472 & 2.08 & 1.31  $\pm$ 0.03 & 230  $\pm$ 7 & 1.97  $\pm$ 0.07 &   \\
 SDSSJ140228.21+632133.5 & 0.205 & 0.481 & 2.65 & 1.35  $\pm$ 0.03 & 285  $\pm$ 7 & 2.24  $\pm$ 0.15 &   \\
 SDSSJ153150.07-010545.7 & 0.160 & 0.744 & 2.73 & 1.71  $\pm$ 0.03 & 273  $\pm$ 9 & 2.07  $\pm$ 0.25 &   \\
 SDSSJ153812.92+581709.8 & 0.143 & 0.531 & 1.45 & 1.00  $\pm$ 0.02 & 236  $\pm$ 5 & 2.27  $\pm$ 0.12 &   \\
 SDSSJ162132.99+393144.6 & 0.245 & 0.602 & 2.30 & 1.27  $\pm$ 0.03 & 255  $\pm$ 10 & 2.02  $\pm$ 0.07 &   \\
 SDSSJ162746.45-005357.6 & 0.208 & 0.524 & 2.02 & 1.22  $\pm$ 0.02 & 262  $\pm$ 10 & 1.93  $\pm$ 0.14 &   \\
 SDSSJ163028.16+452036.3 & 0.248 & 0.793 & 2.01 & 1.79  $\pm$ 0.04 & 280  $\pm$ 7 & 1.92  $\pm$ 0.09 &   \\
 SDSSJ230321.72+142217.9 & 0.155 & 0.517 & 3.46 & 1.62  $\pm$ 0.03 & 259  $\pm$ 9 & 1.94  $\pm$ 0.08 &   \\
\hline
 B1608+656 & 0.630 & 1.394 & 0.59 & 0.81$\pm$0.05 & 266$\pm$16 & 2.08$\pm$0.11 \\
 DES0408-5354 & 0.597 & 2.375 & 1.94 & 1.92$\pm$0.10 & 222$\pm$26 & 1.90$\pm$0.10 \\
 HE0435-1223 & 0.455 & 1.693 & 1.80 & 1.22$\pm$0.08 & 214$\pm$15 & 1.93$\pm$0.10 \\
 PG1115+080 & 0.311 & 1.722 & 0.45 & 1.08$\pm$0.06 & 292$\pm$26 & 2.17$\pm$0.12 \\
 RXJ1131-1231 & 0.295 & 0.654 & 1.91 & 1.63$\pm$0.08 & 313$\pm$20 & 1.95$\pm$0.11 \\
 SDSS1206+4332 & 0.745 & 1.789 & 0.29 & 1.25$\pm$0.06 & 310$\pm$32 & 1.95$\pm$0.11 \\
 WFI2033-4723 & 0.658 & 1.662 & 1.97 & 0.94$\pm$0.05 & 250$\pm$19 & 1.95$\pm$0.10 \\
 WGD2038-4008 & 0.228 & 0.777 & 2.22 & 1.38$\pm$0.07 & 286$\pm$19 & 2.3$\pm$0.12 \\
\bottomrule
\end{tabular} }
\end{threeparttable}
\end{center}
\label{tab:data}
\tablefoot{For each dataset (S2LS, SLACS / KCWI, TDCOSMO), we report object name, deflector and source redshift, effective radius, Einstein radius, stellar velocity dispersion corrected to an equivalent aperture of radius equal to half the effective radius, and slope of the mass density profile as inferred from lensing. References are: SL2S \citep{William_2025_dinos_2, Sonnenfeld_2013a}; SLACS\citep{Auger_2010,T24,Knabel24}; TDCOSMO \citep{Milestone25, birrer20}}
\end{table*}
\end{appendix}

%

%


\end{document}